\pdfoutput=1
\documentclass[11pt]{article}

\newcommand{\newc}{\newcommand}
\def\eq$#1${\begin{equation}#1\end{equation}}
\def\gat$#1${\begin{gather}#1\end{gather}}
\def\bal$#1${\begin{align}#1\end{align}}
\def\eqarr$#1${\begin{eqnarray}#1\end{eqnarray}}
\newc{\pa}{\partial}
\newc{\alp}{\alpha}
\newc{\gam}{\gamma}
\newc{\Gam}{\Gamma}
\newc{\del}{\delta}
\newc{\eps}{\epsilon}
\newc{\lam}{\lambda}
\newc{\sig}{\sigma}
\newc{\ups}{\upsilon}
\newc{\ome}{\omega}
\newc{\nonum}{\nonumber}
\newc{\vph}{\varphi}
\newc{\tx}{\tilde{x}}
\newc{\paper}[2]{\enquote{\textit{#1}} \cite{#2}}
\newc{\erf}{\text{erf}}
\newcommand{\myref}[2]{{\color{MyDarkBlue}\ref{#1}(\subref{#2})}} 

\usepackage{graphicx,rotating,colortbl}
\usepackage{jcappub}
\usepackage{amsmath}
\usepackage{amssymb}
\usepackage{braket}
\usepackage{graphicx}
\usepackage{hyperref}
\usepackage{mathrsfs}
\usepackage{subcaption}
\usepackage{empheq}
\usepackage{appendix}
\usepackage{natbib}
\usepackage[dvipsnames]{xcolor}
\usepackage{float}\usepackage[top=1in, bottom=0.4in, left=0.8in, right=0.8in, includefoot]{geometry} 
\usepackage[autostyle]{csquotes}
\usepackage{color,soul}
\usepackage{mathtools}
\usepackage{multirow}
\usepackage{tabu}
\usepackage{fixmath}

\usepackage{ulem} 

\usepackage{titlesec}

\titlespacing\section{0pt}{12pt plus 4pt minus 2pt}{5pt plus 2pt minus 2pt}
\titlespacing\subsection{0pt}{12pt plus 4pt minus 2pt}{12pt plus 2pt minus 2pt}

\definecolor{azure(colorwheel)}{rgb}{0.0, 0.5, 1.0}
\definecolor{DarkViolet}{RGB}{148,0,211}
\definecolor{MyDarkBlue}{rgb}{0,0.1,0.7}
\definecolor{DarkBlue}{RGB}{0,0,153}
\definecolor{amber}{rgb}{1.0, 0.49, 0.0}
\definecolor{amaranth}{rgb}{0.9, 0.17, 0.31}
\definecolor{nicered}{rgb}{0.7,0.1,0.1}
\definecolor{brown}{rgb}{0.5,0.1,0.1}
\definecolor{nicegreen}{rgb}{0.0,0.3,0.0}
\definecolor{tealgreen}{rgb}{0.0, 0.51, 0.5}
\hypersetup{colorlinks,bookmarksopen,
bookmarksnumbered,
citecolor={nicered},
linkcolor={MyDarkBlue},
urlcolor={tealgreen},
pdfstartview=FitH}

\graphicspath{{Figs/}}
\newcommand{\comm}[1]{}

\newc{\teo}[1]{\textcolor{azure(colorwheel)}{#1}} 
\newc{\com}[1]{\textcolor{amaranth}{#1}} 
\newc{\bako}[1]{\textcolor{DarkViolet}{#1}} 
\newc{\corr}[1]{\textcolor{red}{#1}} 

\usepackage{scalerel}
\usepackage{tikz}
\usetikzlibrary{svg.path}
\definecolor{orcidlogocol}{HTML}{A6CE39}
\tikzset{
  orcidlogo/.pic={
    \fill[orcidlogocol] svg{M256,128c0,70.7-57.3,128-128,128C57.3,256,0,198.7,0,128C0,57.3,57.3,0,128,0C198.7,0,256,57.3,256,128z};
    \fill[white] svg{M86.3,186.2H70.9V79.1h15.4v48.4V186.2z}
                 svg{M108.9,79.1h41.6c39.6,0,57,28.3,57,53.6c0,27.5-21.5,53.6-56.8,53.6h-41.8V79.1z M124.3,172.4h24.5c34.9,0,42.9-26.5,42.9-39.7c0-21.5-13.7-39.7-43.7-39.7h-23.7V172.4z}
                 svg{M88.7,56.8c0,5.5-4.5,10.1-10.1,10.1c-5.6,0-10.1-4.6-10.1-10.1c0-5.6,4.5-10.1,10.1-10.1C84.2,46.7,88.7,51.3,88.7,56.8z};}}
\newcommand\orcid[1]{\href{https://orcid.org/#1}{\mbox{\scalerel*{
\begin{tikzpicture}[yscale=-1,transform shape]
\pic{orcidlogo};
\end{tikzpicture}
}{|}}}}

 \vspace*{0cm}

\title{Novel exact ultra-compact and ultra-sparse hairy black holes emanating from regular and phantom scalar fields}

\author[a,b]{Athanasios Bakopoulos\,\orcid{0000-0002-3012-6144}}
\author{and}
\author[a]{Theodoros Nakas\,\orcid{0000-0002-3522-5803}}

\emailAdd{a.bakop@uoi.gr}
\emailAdd{theodoros.nakas@gmail.com}

\affiliation[a]{Physics Division, School of Applied Mathematical and Physical Sciences,
National Technical University of Athens, Zografou Campus, Zografou, GR-15780, Greece\\}
\affiliation[b]{Division of Applied Analysis, Department of Mathematics,
University of Patras, Rio Patras GR-26504, Greece\\}

\abstract{In the framework of a simple gravitational theory that contains a scalar field minimally coupled to gravity, we investigate the emergence of analytic black-hole solutions with non-trivial scalar hair of secondary type. Although it is possible for one to obtain asymptotically (A)dS solutions using our setup, in the context of the present work, we are solely interested in asymptotically flat solutions. At first, we study the properties of static and spherically symmetric black-hole solutions emanating from both regular and phantom scalar fields. We find that the regular-scalar-field-induced solutions are solutions describing ultra-compact black holes, while the phantom scalar fields generate ultra-sparse black-hole solutions. The latter are black holes that can be potentially of very low density since, contrary to ultra-compact ones, their horizon radius is always greater than the horizon radius of the corresponding Schwarzschild black hole of the same mass. Then, we generalize the above static solutions to slowly rotating ones and compute their angular velocities explicitly. Finally, the study of the axial perturbations of the derived solutions takes place, in which we show that there is always a region in the parameter space of the free parameters of our theory that allows the existence of both ultra-compact and ultra-sparse black holes.}

\begin{document}

\maketitle
\vspace*{0.8cm}


\section{Introduction}
\label{Sec: intro}

In this day and age, it is considered common knowledge that the General Theory of Relativity (GR) is merely an effective theory and is therefore expected to break down at some energy scale.
The Hierarchy Problem, the unknown nature of Dark Energy and Dark Matter, as well as the incompetence of GR to address by itself an inflationary era in our universe, are only some of the most notable unresolved problems that lead us to examine modified gravitational theories.
Throughout the years, there have been formulated various modified gravitational theories, however, the most basic and extensively studied ones are the scalar-tensor theories.
These theories provide an additional scalar degree of freedom introduced via the existence of a scalar field which is either minimally or non-minimally coupled with gravity.
Owing to the fact that the vast majority of modified gravitational theories reduce to scalar-tensor theories in a particular limit, scalar-tensor theories render a very fertile framework for developing new ideas and investigating new spacetime geometries.
In addition to the above, the detection of the Higgs boson in 2012 \cite{Aad:2012tfa,Chatrchyan:2012ufa}, has made physicists widely acknowledge the existence of scalar fields in nature.

The credibility of Modified Gravitational theories, which are typically formulated as cosmological models, relies significantly on their ability to incorporate local solutions such as black holes, neutron stars, or stars.
If the theory cannot produce astrophysical realistic local solutions, then it cannot be considered a viable model.
Although the search for new black hole solutions in scalar-tensor theories was prematurely curtailed due to the formulation of No-Scalar Hair theorems \cite{NH-scalar1, NH-scalar2,Bekenstein}, it was shortly shown that these theorems can be circumvented. 
As a result, a plethora of hairy black holes have then appeared in the literature.
A partial list of asymptotically flat solutions can be found in Refs. \cite{Torii:1993vm, Bechmann:1995sa, Dennhardt:1996cz, Nucamendi:1995ex, Gubser:2005ih, Bronnikov:2005gm, Nikonov_2008, Anabalon:2012ih, Anabalon:2013qua, Kleihaus:2013tba, Babichev:2013cya, Sotiriou:2014pfa, Herdeiro:2014goa, Charmousis:2014zaa, Astorino:2014mda, Cadoni:2015gfa, Herdeiro:2015gia, Kleihaus:2015iea, Kanti:2019upz, Bakopoulos:2019fbx, Tahamtan:2015sra, Tolley:2015ywa, Hod:2017kpt, Herdeiro:2016tmi, Ni:2016rhz, Benkel:2016rlz, Sanchis-Gual:2016tcm, Heisenberg:2017xda, Antoniou:2017acq,  Doneva:2017bvd, Silva:2017uqg, Antoniou:2017hxj, Herdeiro:2018wvd, Pacilio:2018gom, Brihaye:2018woc, Brihaye:2018grv, Wang:2018xhw, Herdeiro:2018daq,Astefanesei:2019mds, VanAelst:2019kku, Hod:2019pmb, Kunz:2019sgn, Cunha:2019dwb, Filippini:2019cqk, Zou:2019ays, Fernandes:2020gay, Santos:2020pmh, Sultana:2020pcc, Hong:2020miv, Astefanesei:2020xvn, Shnir:2020hau, Hod:2020jjy, Ovalle:2020kpd, Delgado:2020hwr, Myung:2020ctt, Brihaye:2021ich,Faraoni:2021nhi,Bakopoulos:2022csr, Karakasis:2021lnq, Karakasis:2021rpn, Karakasis:2021ttn, Karakasis:2022fep, Karakasis:2022xzm, Theodosopoulos:2023mtk, Chatzifotis:2022ene, Babichev:2023dhs, Fernandes:2021dsb}, while for asymptotically (A)dS$_4$ solutions, the reader is referred to \cite{Martinez:2004nb, Martinez:2006an, Anabalon:2012ta, Charmousis:2015aya, Babichev:2015rva, Fan:2015oca, Perapechka:2016cof, Bakopoulos:2018nui, BenAchour:2018dap, Brihaye:2019gla, Guo:2021zed, Babichev:2023rhn}.
Hairy black holes constitute a subject of intense study over the past decades, since they have observable effects, such as the emission of gravitational waves, black-hole shadows, modifications in their accretion disks, etc.
Thus, they might provide insights into the fundamental nature of gravity.
These solutions have been studied in both the classical and quantum regimes, and have been found to exhibit a variety of interesting phenomena, such as spontaneous scalarization  \cite{Antoniou:2017acq, Doneva:2017bvd, Silva:2017uqg, Bahamonde:2022chq, Staykov:2022uwq, Doneva:2020nbb, Doneva:2021dcc, Doneva:2021dqn, Doneva:2022ewd, Doneva:2022yqu, Antoniou:2022agj, Antoniou:2020nax, Herdeiro:2020wei, Dima:2020yac, Ventagli:2020rnx, Andreou:2019ikc, Macedo:2019sem, Tang:2020sjs} and bifurcations \cite{Dias:2010ma, Zou:2019ays, Zou:2020zxq, Guo:2021ere}.
It is important to note at this point, that the sole existence of black holes does not guarantee that these objects are serious candidates for astrophysical objects since astrophysical black holes must be both stable and rotating. 
To this end, it is crucial to investigate the existence of stable and rotating black-hole solutions in the context of scalar-tensor theories.

In conjunction with the above, scalar-tensor theories possess an additional advantage, which is the fact that they are able to generate local solutions that are prohibited in the context of GR altogether, such solutions are wormholes and even solitonic particle-like solutions.
In General Relativity, wormholes necessitate the presence of exotic matter near their throat \cite{Morris:1988cz}.
Likewise, in electrovacuum, solitonic solutions were proven to be unstable \cite{Wheeler:1955zz, Misner:1957mt}.  
However, in the framework of scalar-tensor theories, such as the Einstein-scalar-Gauss-Bonnet and beyond Horndeski gravity, it has been demonstrated that real scalar fields may support regular wormhole solutions \cite{Kanti:2011jz, Kanti:2011yv, Antoniou:2019awm, Chatzifotis:2022mob, Chatzifotis:2021hpg, Babichev:2022awg, Bakopoulos:2022gdv, Bakopoulos:2021liw}.
Additionally, numerous stable solitonic solutions have been discovered within the context of scalar-tensor theories
\cite{Fisher:1948yn, Janis:1968zz, Wyman:1981bd, Agnese:1985xj, Roberts:1989sk, Kleihaus:2019rbg, Herdeiro:2019iwl, Kleihaus:2020qwo, Baake:2021jzv, Chatzifotis:2022ubq}. 
It is worth noting that scalar-tensor theories also predict the existence of ultra-compact black holes \cite{Bakopoulos:2020dfg, Bakopoulos:2021dry}.
These are local solutions with a horizon radius always smaller than that of the corresponding GR black holes of the same mass. 
Finally, as we are about to see in the present work, in the context of a very simple theory with the existence of a scalar field minimally coupled with gravity, it is possible to obtain besides ultra-compact black-hole solutions, solutions of ultra-sparse black holes.
The latter solutions emanate from a phantom scalar field and describe local objects which, in contrast to the ultra-compact black holes, are less dense than the corresponding GR black holes of equal mass.
All the previously mentioned solutions could serve as precise models for local astronomical objects observed in the universe, such as the X-ray transient GROJ0422+32 \cite{compact} (also refer to \cite{LIGOScientific:2018mvr, Abbott:2020uma, LIGOScientific:2020ibl}).

The present study is focused on a simple action functional that comprises a minimally coupled scalar field with both kinetic and potential terms, in addition to Einstein's gravity. 
This theory is a member of the Horndeski class,\,\footnote{It is a Horndeski theory with $G_2=X+V(\Phi)$, $G_4=1$ and $G_3=G_5=0$.} and is conformally equivalent to both $f(R)$ and Brans-Dicke theories. 
Also, the theory has been widely used in Cosmology, as it offers accurate models for dark energy and inflation. 
Regarding the local solutions, the theory was employed early on for the construction of wormhole solutions, with the Ellis wormhole \cite{Ellis:1973yv, Ellis:1979bh, Bronnikov:1973fh} being a characteristic example that is supported by phantom fields. 
However, due to the no-scalar hair theorems \cite{NH-scalar1, NH-scalar2,Bekenstein}, it is not possible to derive black hole solutions for a broad range of potentials. 
 Specifically, black-hole solutions emanating from a regular scalar field can only be obtained for negative definite potentials, i.e., $V(\Phi) < 0$, while---as we are about to see in Sec. \ref{Sec: theory}---similar solutions emanating from a phantom scalar field necessitate a positive-definite potential. 
 To this end, it is crucial to identify the potentials that lead to analytic solutions for black holes. 
 Several works \cite{Bechmann:1995sa, Dennhardt:1996cz, Nucamendi:1995ex, Anabalon:2012ih, Cadoni:2015gfa} have been devoted to this direction in recent years, and they commonly employ the ``scalar-potential engineering" method, in which the form of the scalar field is specified, and the potential is determined by solving the field equations. 
 In this work, we consider several exact/analytic black-hole solutions and investigate their properties in depth. 
 We demonstrate that the solutions can indeed describe both ultra-compact and ultra-sparse black holes, and then these solutions are generalized into slowly rotating black holes. We also compare the angular velocities of these solutions with those of corresponding slowly rotating Schwarzschild black holes of the same mass. 
 Finally, in order to evaluate whether these solutions could represent astrophysical objects, we examine their thermodynamic stability and their stability under spacetime perturbations.

The structure of this paper is as follows: in Section \ref{Sec: theory}, we introduce our four-dimensional field theory, and then we derive the black-hole geometry by solving the field equations. 
We then examine the geometric and thermodynamic properties of our solutions, and we also investigate whether the field theory accompanying our solutions fulfills the conditions for the evasion of the no-scalar hair theorem \cite{Bekenstein}.
In Section \ref{Sec: rot-BHs}, we generalize the static solutions into slowly rotating ones by using the slow-rotation approximation, which was first proposed by Hartle \cite{Hartle:1967he} in the framework of General Relativity and generalized by Pani and Cardoso for scalar-tensor theories \cite{Pani:2009wy}.
In Section \ref{Sec: stab}, we study the stability of our black hole solutions under axial perturbations. 
Finally, we summarize our analysis and discuss our results in Section \ref{Sec: concl}.


\section{Theoretical framework and static black-hole solutions}
\label{Sec: theory}

The class of theories known as Einstein-scalar-Gauss-Bonnet theories represents a distinctive yet highly comprehensive group of generalized gravitational theories. These theories, in addition to the conventional Einstein term, incorporate a scalar field and the quadratic Gauss-Bonnet term. Despite their inherent simplicity, they possess an extensive range of complexities and intricacies. The action of the theory takes the form 
\eq$\label{actiongb}
\mathcal{S}=\frac{1}{16\pi}\int d^4x\sqrt{-g}\bigg[R-\frac{1}{2}(\pa \Phi)^2 -V(\Phi)+ \alpha f(\Phi) R^2_{GB}\bigg]\,.$
The theory contains the Einstein-Hilbert term $R\equiv g^{\mu\nu}R_{\mu\nu}$, and a scalar field $\Phi$ 
non-minimally coupled with the gravitational field. In the above action and from this time forth, we use the notation $(\pa \Phi)^2\equiv \pa^\mu\Phi\pa_\mu\Phi$. Also, the Gauss-Bonnet term is defined as 
\begin{equation}
    R^2_{GB}=R^{\mu\nu\rho\sigma}R_{\mu\nu\rho\sigma}-4R^{\mu\nu}R_{\mu\nu}+R^2.
\end{equation}
It can be shown that due to the presence of the Gauss-Bonnet term the above theory violates the weak energy condition near the horizon of a black hole and therefore lead to the evasion of the no-scalar-hair theorem \cite{Antoniou:2017acq}. In the work \cite{Bakopoulos:2020dfg}, it was demonstrated that the theory allows for asymptotically flat black hole solutions through the utilization of a negative-definite potential. When the dimensionless coupling constant $\alpha/r_h^2$ is small, we may treat the Gauss-Bonnet coupling as a small interaction. The $\mathcal{O}(1)$ terms in the expansion of $\alpha/r_h^2$ can be described by the following action functional
\eq$\label{action}
\mathcal{S}=\frac{1}{16\pi}\int d^4x\sqrt{-g}\bigg[R-\frac{1}{2}(\pa \Phi)^2-V(\Phi)\bigg]\,.$
The form of the background solution is contingent upon the nature of the potential employed. In the case where the potential is positive definite, the background solution would align with the Schwarzschild black hole, as governed by the no-scalar-hair theorem. Conversely, in the context of the Einstein scalar Gauss-Bonnet theory, where the potential is negative, the background solution would manifest as a distinctive and non-trivial solution. Hence, it becomes crucial to ascertain the solutions arising from the action  \eqref{action}, even when the potential is negative definite. This significance stems from the subsequent utilization of the Gauss-Bonnet term, which, when allowed to back react on the background solution, enables the attainment of a new viable and realistic astrophysical solution.  It is worth noting that the action \eqref{action} serves as the foundation for numerous generalized theories. As a result, the library of solutions generated can serve as a background not only for the Gauss-Bonnet interaction but also for a wider range of interactions, further enhancing its applicability and versatility.
 
 From the variation of the action \ref{action} with respect to the metric tensor $g^{\mu\nu}$ we find the following tensorial equation 
\begin{equation}
    G^{\mu}{}_\nu=T^{(\Phi)\mu}{}_\nu\,,\label{gr-eqs}
\end{equation}
where $T^{(\Phi)\mu}{}_\nu$ is the effective stress-energy tensor associated with the presence of the scalar field $\Phi$ and is defined as
\eq$\label{stress-ten}
T^{(\Phi)\mu}{}_\nu\equiv \frac{1}{2}\pa^\mu\Phi\pa_\nu\Phi-\frac{1}{2}\del^{\mu}{}_\nu\left[\frac{(\pa \Phi)^2}{2}+V(\Phi)\right]\,.$
By varying the action with respect to the scalar field $\Phi$ we obtain the following equation of motion:
 \begin{equation}
  \nabla^\lam \nabla_\lam \Phi-\pa_\Phi V =0\,.  \label{sc-eq}
\end{equation} 
Within the scope of this study, our focus lies on deriving asymptotically flat black-hole solutions that possess scalar hair.
Therefore,  we consider the following ansatz for the line-element
\eq$\label{metr-ans}
ds^2=-e^{A(r)}\, B(r)\, dt^2+\frac{dr^2}{B(r)}+r^2\big(d\theta^2+\sin^2\theta\, d\varphi^2\big)\,,
\hspace{1em} B(r)\equiv 1-\frac{2m(r)}{r}\,.$
It is also reasonable to presume that the scalar field  depends only on the radial coordinate, namely $\Phi=\Phi(r)$. 
Using now the tensorial equation \eqref{gr-eqs}, we obtain the following independent equations
\eq$\label{eq1}
A'(r)=\frac{r}{2} \left[\Phi'(r)\right]^2\,,$
\eq$\label{eq2}
B''(r) + \frac{3}{2}\, A'(r) B'(r) + \left\{ A''(r) + \frac{A'(r)}{r} + \frac{\left[A'(r)\right]^2}{2} - \frac{2}{r^2} \right\} B(r) = 
-\frac{2}{r^2} \,,$
\eq$\label{eq3}
V(\Phi) = \frac{2}{r^2} -\frac{2}{r}\, A'(r) B(r)-\frac{2B(r)}{r^2} +\frac{1}{2}\left[\Phi'(r)\right]^2 B(r) -\frac{2B'(r)}{r}\,. $
In the above, the prime symbol is used to represent differentiation with respect to the radial coordinate, $r$. 
Furthermore, it is important to acknowledge that the scalar-field equation \eqref{sc-eq}  is not an independent one, but instead is derived from the three differential equations previously mentioned. An explicit proof of the previous assessment can be found in Appendix \ref{App: sc-eq}.
At this point, it is apparent that to tackle the system of differential equations \eqref{eq1}-\eqref{eq3} which encompasses four unknown functions, we must initially specify a particular expression for one of the functions involved.

Contrary to what we have done in \cite{Bakopoulos:2021dry}, where we assumed a Coulombic form for 
the scalar field $\Phi(r)$, here we assume an expression for the function $A(r)$, namely
\eq$\label{eq: A-r}
A(r) = - \xi\, \ln \left( \frac{1+r^2/q^2}{r^2/q^2} \right)\,.$
In the above, $\xi$ is a dimensionless constant that takes values in real numbers, while $q$ is a physical parameter with length units which is strictly positive. 
Using eq. \eqref{eq1} we find that
\eq$\label{eq: phi-r}
\Phi(r) = 2 \sqrt{\xi}\, \ln \left( \frac{1+\sqrt{1+r^2/q^2}}{r/q} \right)\,.$
Notice that for $\xi>0$ we have a real-valued scalar field, while for $\xi<0$ the scalar field becomes purely imaginary.
In the latter case, by performing a scalar field redefinition of the form $\Phi=i \tilde{\Phi}$, one is led to the action
\eq$\label{eq: phan-act}
\mathcal{S}=\frac{1}{16\pi}\int d^4x\sqrt{-g}\bigg[R+\frac{1}{2}\left(\pa \tilde{\Phi}\right)^2-V(\tilde{\Phi})\bigg]\,,$
with $\tilde{\Phi}$ being a phantom scalar field---as it is often called---due to the fact that its kinetic term comes with a different sign.
It is evident from eq. \eqref{eq3} that the scalar potential $V(\tilde{\Phi})\in \mathbb{R}$ even in the case of phantom field.
Consequently, by deciding on the sign of $\xi$, one may choose between normal\,\footnote{Solutions which originate from a real-valued scalar field ($\xi>0$) will be called \textit{normal} throughout this work to be distinguished from the phantom ones ($\xi<0$). We avoid calling them \textit{regular} since in the literature, regular are usually called the black holes which do not contain a singularity.} or phantom solutions.
Finally, the trivial case $\xi=0$ leads to the well-known Schwarzschild solution.

Having determined the scalar field from \eqref{eq1}, we are now pertaining to \eqref{eq2} in order to determine the mass function $m(r)$ or equivalently the function $B(r)$.
Substituting the expression for $A(r)$, the differential equation \eqref{eq2} takes the form
\eq$\label{eq2-new}
B''(r) + \frac{3\xi q^2}{r\left(q^2+r^2\right)}\, B'(r) + 
2\, \frac{q^4(\xi^2-1)-2q^2r^2(\xi+1)-r^4}{r^2\left(q^2+r^2\right)^2}\, B(r) = -\frac{2}{r^2}\,.$
By solving the preceding differential equation one should obtain
\bal$\label{eq: B-r}
B(r) = &\,\left( r/q\right)^{2(1-\xi)} \left(1+r^2/q^2\right)^\xi \left[ C_1 + 2 \int (r/q)^{3\xi-2}\left(1+r^2/q^2\right)^{-3\xi/2} H(r)\, d(r/q)\right] \nonum\\
& + H(r) \left[ C_2 -2 \int (r/q)^\xi \left( 1+r^2/q^2 \right)^{-\xi/2} \, d(r/q) \right] \,,$
where
\eq$\label{eq: H-r}
H(r) \equiv \left\{ \begin{array}{cr}
\displaystyle{\left(r/q\right)^{2(1-\xi)}\left(1+r^2/q^2\right)^\xi \int \left( r/q \right)^{\xi-4} \left( 1+r^2/q^2  \right)^{-\xi/2}\ d\left( r/q \right)}\,, & \hspace{1em}\xi\in\mathbb{Z}\\[5mm]
\displaystyle{ \left( r/q \right)^{-(1+\xi)} \frac{\left(1+r^2/q^2\right)^\xi}{\xi-3}\, \,_2F_1\left(\frac{\xi}{2},\frac{\xi-3}{2};\frac{\xi-1}{2};-\frac{r^2}{q^2}\right)}\,, & \xi\notin\mathbb{Z}
\end{array}\right\}\,.$
For $|z| < 1$ the hypergeometric function $\,_2F_1(a,b;c;z)$ can be determined via the following convergent infinite 
series (see \cite{abramowitz+stegun})
\eq$\label{eq: hyper-exp}
\,_2F_1(a,b;c;z) = \,_2F_1(b,a;c;z) = \sum_{n=0}^\infty \frac{a^{(n)} b^{(n)}}{c^{(n)}} \frac{z^n}{n!}\,.$
The symbols of the form $d^{(n)}$ are rising Pochhammer symbols which are given by the following relation
\eq$\label{eq: poch}
d^{(n)} = \frac{\Gamma(d+n)}{\Gamma(d)} = \left\{ \begin{array}{cr}
d(d+1)\cdots (d+n-1)\,, & \hspace{0.5em} n>0 \\[2mm]
1\,, & \hspace{0.5em} n=0 
\end{array} \right\} \,.$
Although the argument $-r^2/q^2$ of the hypergeometric function in \eqref{eq: H-r} does not respect the aforementioned condition, by applying a Pfaff transformation of the form
\gat$\label{eq: Pfaff}
\,_2F_1 (a,b;c;z) = (1-z)^{-a}\, \,_2F_1 \left( a, c-b; c; \frac{z}{z-1} \right) \,,$
we can make the argument of the resulting hypergeometric function smaller than unity. Then, the expansion \eqref{eq: hyper-exp} can be directly applied.

Pertaining now to the solution \eqref{eq: B-r}, it might seem that it was given by a deus ex machina, however, in Appendix \ref{App: DE}, we provide a step-by-step treatment of differential equation \eqref{eq2-new}.
It is important to clarify that for $\xi\in \mathbb{Z}$ the integrals in eqs. \eqref{eq: B-r} and \eqref{eq: H-r} can be solved in terms of known functions, and thus $B(r)$ is determined analytically.
On the contrary, for non-integer values of $\xi$ one may use the expression
\bal$\label{eq: B-r-xi-non-int}
B(r) = &\, C_1\, (r/q)^{2(1-\xi)} \left( 1+r^2/q^2 \right)^\xi + \frac{C_2}{\xi-3}\, (r/q)^{-(1+\xi)} \left( 1+r^2/q^2 \right)^\xi 
\,_2F_1 \left( \frac{\xi}{2}, \frac{\xi-3}{2}; \frac{\xi-1}{2}; -\frac{r^2}{q^2} \right) \nonum\\
& + \frac{2 \left( 1+r^2/q^2 \right)^\xi}{\xi-3} \Bigg\{
 \left(\frac{r}{q}\right)^{2(1-\xi)} \int \,_2F_1 \left( \frac{\xi}{2}, \frac{\xi-3}{2}; \frac{\xi-1}{2}; -\frac{r^2}{q^2} \right) \left(\frac{r}{q}\right)^{2\xi-3} 
\left( 1+\frac{r^2}{q^2} \right)^{-\xi/2} d\left(\frac{r}{q}\right)\nonum\\
& - \frac{1}{\xi+1}\, \,_2F_1 \left( \frac{\xi}{2}, \frac{\xi-3}{2}; \frac{\xi-1}{2}; -\frac{r^2}{q^2} \right)
\,_2F_1 \left( \frac{\xi}{2}, \frac{\xi+1}{2}; \frac{\xi+3}{2}; -\frac{r^2}{q^2} \right) \Bigg\} \,.$
However, since our goal is to keep the analysis as lucid as possible, we will restrain ourselves to integer values of $\xi$.

In what follows, we are going to study one normal and one phantom solution with $\xi=5$ and $\xi=-2$, respectively. 
For each positive value of $\xi$ one gets a different normal solution, while for each negative value of $\xi$ a different phantom solution appears.
The aforementioned choices for the values of $\xi$ have been made because in both cases, the function $B(r)$ has a fairly simple expression. 
Obviously, one is free to choose differently.\,\footnote{In Appendix \ref{App: Add-sol}, one may find two additional asymptotically flat solutions.}
So, for $\xi=5$, \eqref{eq: B-r} leads to
\eq$\label{eq: B-real}
B(r)=\left(1+\frac{r^2}{q^2}\right)^2\left[\frac{q^4}{r^4} + \frac{3q^6}{r^6} + \frac{17}{9} \frac{q^8}{r^8} - \frac{2M}{q} \frac{q^8}{r^8} \left(1+\frac{r^2}{q^2} \right)^{3/2}\, \right]\,,$
while for $\xi=-2$, the same expression results in
\eq$\label{eq: B-phant}
B_\mathfrak{p}(r)=\left(1+\frac{r^2}{q^2}\right)^{-2} \left[ \frac{r^4}{q^4} - \frac{2M}{q}\frac{r}{q} \left(\frac{3}{5}+\frac{r^2}{q^2} \right) - \frac{1}{3} \right]\,.$
The subscript ``$\mathfrak{p}$'' denotes that we are referring to the phantom solution.
This notation will be henceforth used to all phantom-related functions.
Note that in both of the above expressions, the integration constants $C_1$ and $C_2$ have been chosen to be $C_1=0$ and $C_2=6M/q$. 
These particular choices were made in order for the spacetime \eqref{metr-ans} to be asymptotically flat.
Allowing $C_1\neq 0$, one could also obtain asymptotically (anti-)de Sitter spacetime geometries. 
However, in the context of the present work, we will solely focus on asymptotically flat solutions.
The last assertion can be verified by noticing that
\eq$ \lim_{r\rightarrow +\infty} e^{A(r)}=1\,,$
while the expansions of $B(r)$ and $B_\mathfrak{p}(r)$, at large values of the radial coordinate $r$, are given by
\eq$\label{eq: B-real-exp}
B(r\gg 1) = 1 - \frac{2M}{r} + \frac{5q^2}{r^2} - \frac{7M}{q}\frac{q^3}{r^3} + \mathcal{O}\left(\frac{1}{r^4} \right)\,,$
and
\eq$\label{eq: B-phant-exp}
B_\mathfrak{p} (r\gg 1) = 1 - \frac{2M}{r} - \frac{2q^2}{r^2} + \frac{14M}{5q}\frac{q^3}{r^3} + \mathcal{O}\left(\frac{1}{r^4} \right)\,.$
Note that in the case of a regular scalar field---eq. \eqref{eq: B-real-exp}---the asymptotic behavior of the spacetime cannot be distinguished from the geometry of a Reissner-Nordstr\"{o}m black hole. 
The difference, though, is that in our scenario the parameter $q$ emanates from the existence of a scalar field rather than a gauge field.
Pertaining now to the phantom solution, eq. \eqref{eq: B-phant-exp}, we notice that there is a minus sign in front of the third term in the r.h.s. of eq. \eqref{eq: B-phant-exp}.
This particular difference is directly related to the phantom nature of the scalar field.

Given the functions that describe the spacetime geometry, namely $A(r)$ and $B(r)$, as well as the expression for the scalar field $\Phi(r)$, it is straightforward to use eq. \eqref{eq3} to specify the form of the scalar potential $V(\Phi)$.
In the case of the regular field, by employing eqs. \eqref{eq: A-r}, \eqref{eq: phi-r}, \eqref{eq: B-real} in \eqref{eq3}, and setting $\xi=5$, one finds that
\bal$\label{eq: V-real}
V(\Phi)&= \frac{\sinh^6\left(\frac{\Phi}{2\sqrt{5}}\right)}{18q^3} \Bigg\{ 121 q - 54M \left| \coth \left(\frac{\Phi}{2\sqrt{5}} \right) \right| 
+ \cosh\left(\frac{2\Phi}{\sqrt{5}} \right) \Bigg[ 17q -18M \left| \coth\left(\frac{\Phi}{2\sqrt{5}} \right) \right| \Bigg]\nonum\\[1mm]
&\hspace{3cm} +6 \cosh\left(\frac{\Phi}{\sqrt{5}} \right) \Bigg[ 17q -12M \left| \coth\left(\frac{\Phi}{2\sqrt{5}} \right) \right| \Bigg]\Bigg\}\,.$
In the case of the phantom field, besides using eqs. \eqref{eq: A-r}, \eqref{eq: phi-r}, \eqref{eq: B-phant}, and $\xi=-2$ in \eqref{eq3}, we also need to perform the redefinition $\Phi=i \tilde{\Phi}$ to convert the action functional \eqref{action} to the action \eqref{eq: phan-act} which encompasses the Lagrangian density of a phantom scalar field. By doing so, one should obtain
\bal$\label{eq: V-phant}
V_\mathfrak{p}(\tilde{\Phi})=\frac{\tanh^5\left(\frac{\tilde{\Phi}}{2\sqrt{2}}\right)}{15q^3\cosh\left(\frac{\tilde{\Phi}}{2\sqrt{2}}\right)} \Bigg[ 48M+45q\sinh\left(\frac{\tilde{\Phi}}{2\sqrt{2}}\right) + 5q \sinh\left(\frac{3\tilde{\Phi}}{2\sqrt{2}}\right) \Bigg]\,.$

In order to gain a deeper understanding of the spacetime geometry \eqref{metr-ans}, it is essential to analyze its causal structure as well as the curvature invariants (the Ricci scalar $R\equiv R^{\mu}{}_{\mu}$, the scalar $\mathcal{R}\equiv R^{\mu\nu} R_{\mu\nu}$, and the Kretschmann scalar $\mathcal{K}\equiv R^{\mu\nu\rho\sigma}R_{\mu\nu\rho\sigma}$) that arise from its line-element. 
In Appendix \ref{App: Curv-inv}, one may find the analytic expressions of all three curvature invariant quantities for both the normal ($\xi=5$) and the phantom ($\xi=-2$) solutions.
Here, for brevity, we only present the formulas describing the Ricci scalars, thus it is
\bal$\label{eq: R-real}
R&=\frac{2q^2}{3\,r^{10}}\left(15\,r^6+100\,q^2 r^4+130\,q^4 r^2+51\,q^6 \right)-\frac{108 M q^7}{3\,r^{10}}\left(1+\frac{r^2}{q^2} \right)^{5/2}\,,\\[2mm]
\label{eq: R-phant}
R_\mathfrak{p}&= \frac{4}{5\, q^4 r^2} \left(1+\frac{r^2}{q^2}\right)^{-3} \big[ (18M-5r)\, r^3 +2\, q^2(3M+5r)\, r + 5q^4 \big]\,.$
By simply observing their formulas, it is evident that at $r=0$ the spacetime becomes singular.
The existence of a spacetime singularity together with the Cosmic Censorship Hypothesis \cite{Penrose:1969pc} constitute a clear indication that the line-element \eqref{metr-ans} describes two families of black-hole solutions; normal ones when $\xi>0$ and phantom ones when $\xi<0$.
To convince ourselves that the spacetime singularity is always hidden behind a horizon, we need to study the causal structure of the spacetime \eqref{metr-ans}.
To this end, we consider radial null trajectories in the background geometry, while by keeping both the polar angle $\theta$ and the azimuthal angle $\varphi$ constants, the relation $ds^2=0$ results in
\eq$\label{eq: null-tr}
\frac{dt}{dr}=\pm\, \frac{e^{-A(r)/2}}{|B(r)|} =\pm \left( \frac{1 + r^2/q^2}{r^2/q^2} \right)^{\xi/2} \frac{1}{|B(r)|} \,.$
At the boundary of the spacetime, where $r\rightarrow +\infty$, both normal and phantom solutions, with $B(r)$ given by \eqref{eq: B-real} and \eqref{eq: B-phant}, respectively, lead to $dt/dr=\pm 1$ as expected. 
This behavior is indeed common for all asymptotically flat spacetimes.
On the other hand, the horizon of the black hole is defined as the region of space where the ratio $dt/dr$ diverges.
In our case, eq. \eqref{eq: null-tr} suggests that the black-hole horizon radius, $r_h$, would be a root of the equation $B(r_h)=0$, or equivalently a root of the equation $g_{tt}(r_h)=0$.

\begin{figure}[t]
    \centering
    \begin{subfigure}[b]{0.49\textwidth}
    \includegraphics[width=1\textwidth]{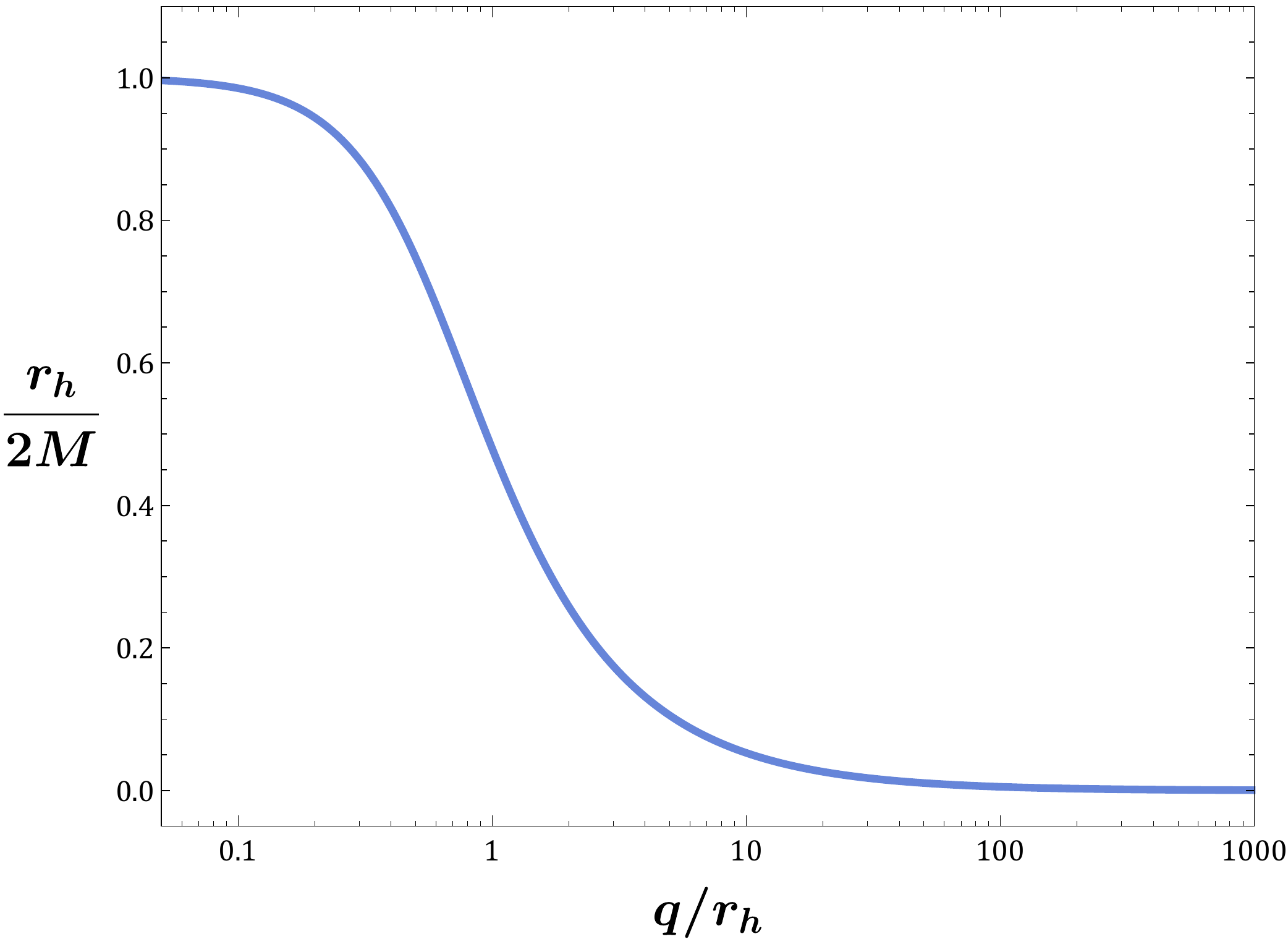}
    \caption{\hspace*{-2.5em}}
    \label{subf: real-rh}
    \end{subfigure}
    \hfill
    \begin{subfigure}[b]{0.485\textwidth}
    \includegraphics[width=1\textwidth]{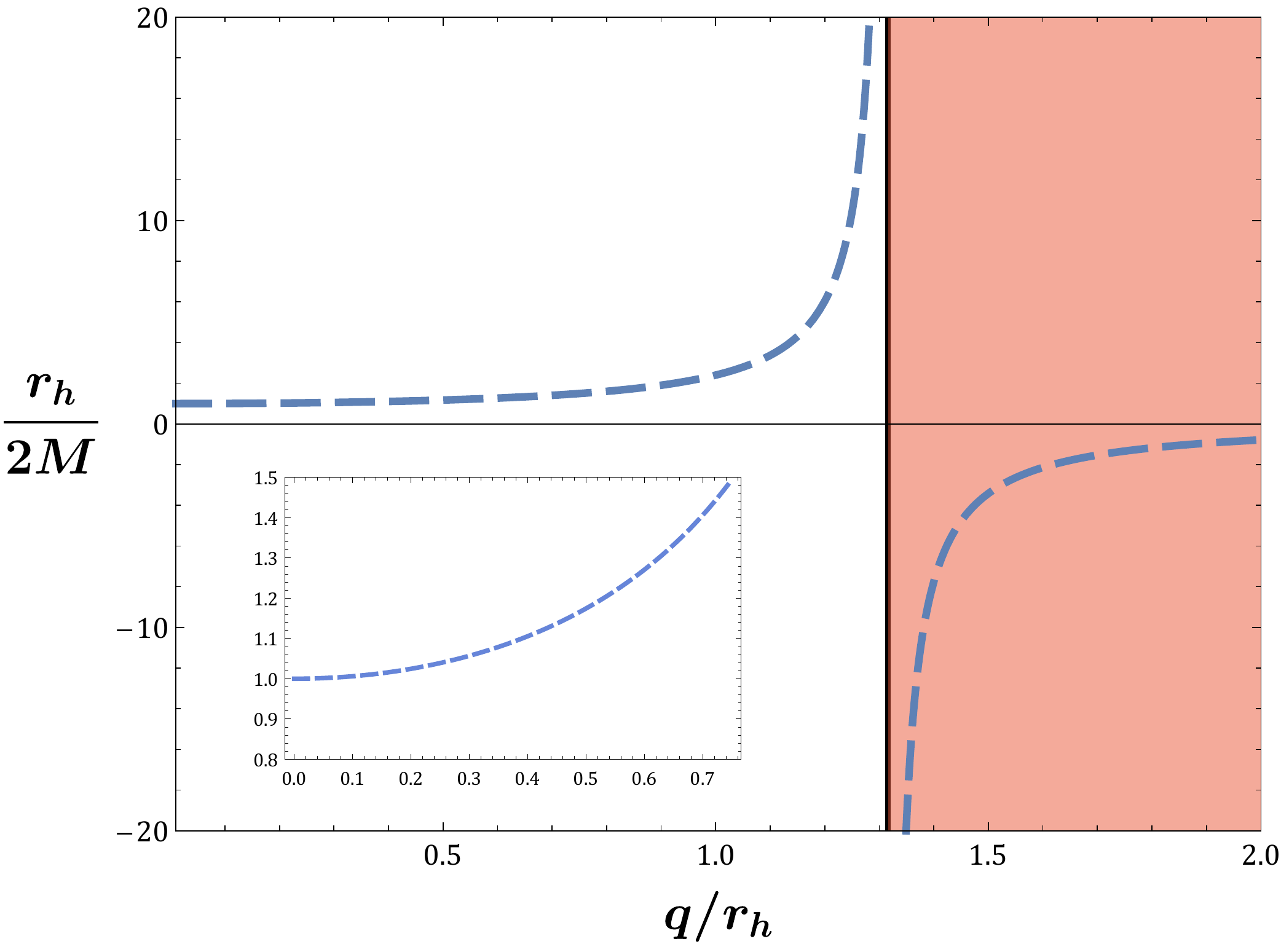}
    \caption{\hspace*{-3em}}
    \label{subf: phant-rh}
    \end{subfigure}
    \caption{The graphs of the ratio $r_h/(2M)$ in terms of $q/r_h$ for (a) a normal scalar field and (b) a phantom one.
    The horizontal axis in (a) is logarithmic.}
    \label{fig: rh-plots}
\end{figure}

With the use of equation $B(r_h)=0$ and the expressions \eqref{eq: B-real} and \eqref{eq: B-phant}, one can specify the relations which link the black-hole mass $M$ with its horizon radius $r_h$ and its scalar hair $q$.
In what follows, we express the ratio $r_h/(2M)$ in terms of the  dimensionless quantity $q/r_h$ proving that the scalar hair is of secondary type. 
For the normal case \eqref{eq: B-real}, one finds that
\eq$\label{eq: rh-real}
\frac{r_h}{2M}=9\left( 1 + \frac{q^2}{r_h^2} \right)^{3/2}\left( 9+27\, \frac{q^2}{r_h^2}+17\, \frac{q^4}{r_h^4}\right)^{-1}\,,$
while for the phantom case \eqref{eq: B-phant}, one obtains
\eq$\label{eq: rh-phant}
\left( \frac{r_h}{2M} \right)_\mathfrak{p} = \left( 1 - \frac{q^4}{3\,r_h^4} \right)^{-1} \left( 1 + \frac{3\,q^2}{5\,r_h^2} \right)\,.$
Figs. \myref{fig: rh-plots}{subf: real-rh} and \myref{fig: rh-plots}{subf: phant-rh} depict the graphs of \eqref{eq: rh-real} and \eqref{eq: rh-phant}, respectively.

The graph presented in Fig. \myref{fig: rh-plots}{subf: real-rh} illustrates that as the value of $q/r_h$ increases, the value of $r_h/(2M)$ decreases. 
It is also obvious that as $q/r_h$ approaches zero, the ratio $r_h/(2M)$ goes to unity and therefore $r_h= 2M$ as in the Schwarzschild case.
This behavior is indeed anticipated given the fact that for $q=0$ the scalar field $\Phi$ vanishes and the theory \eqref{action} reduces to the Einstein-Hilbert action, which by its turn leads to the Schwarzschild solution.
On the other hand, for $q/r_h\geq 1$, the condition $B(r_h)=0$ leads to the formation of \textit{ultra-compact black holes}, given that in this particular region of the parameter space, $r_h/(2M)< 0.5$.
Loosely speaking, an ultra-compact black hole could be defined as a black hole that has a significantly smaller horizon radius compared to the radius of a Schwarzschild black hole with the same mass.
However, for values of $q/r_h$ greater than 10, the horizon radii of the associated ultra-compact black holes become incredibly small, raising concerns about the naturalness of this region in the parameter space. 

Notably, in the phantom case, we observe a different yet intriguing behavior.
The graph in Fig. \myref{fig: rh-plots}{subf: phant-rh} demonstrates that initially, as the value of $q/r_h$ increases, so does the value of $r_h/(2M)$.
However, by further increasing the value of $q/r_h$, we can observe from both the graph and the eq. \eqref{eq: rh-phant} that at $q/r_h=\sqrt[4]{3}$, the ratio $r_h/(2M)$ diverges and then becomes negative.
As both $r_h$ and $M$ are positive-definite quantities, this implies that the region in the parameter space where $q/r_h\geq \sqrt[4]{3}$ is not physically valid.
The fact that $r_h>2M$ in the region $0<q/r_h<\sqrt[4]{3}$, merely indicates that, unlike ultra-compact solutions, phantom scalar fields result in black holes which are less dense compared to the corresponding Schwarzschild black holes of the same mass $M$.
Henceforth, black-hole solutions for which the ratio $r_h/(2M)$ becomes substantially greater than unity will be called \textit{ultra-sparse black holes}.
Note, as an example, that for $q/r_h=1$, the resulting phantom solution has $r_h/(2M)\approx 2.400$ which means that its volume is more than thirteen times greater than the volume of the corresponding Schwarzschild black hole of mass $M$.
The preceding discussion sheds light on a quite interesting characteristic of the black-hole solutions which emanate from the theory \eqref{action}.
This is the fact that the sole nature of the scalar field, namely whether it is phantom or not, decides the density of the resulting solutions.
Finally, as in the case of ultra-compact solutions, the extreme cases of ultra-sparse black holes, namely those where $q/r_h\rightarrow \sqrt[4]{3}$ and the ratio $r_h/(2M)$ blows up, are not expected to be physically plausible.
Future observational measurements from missions such as LIGO-VIRGO, the Event Horizon Telescope, or other astrophysical experiments will hopefully establish strict bounds on the mass and size of compact objects \cite{LIGOScientific:2020ibl, Abbott:2020uma, LIGOScientific:2018mvr}. 

It is well-known that the entropy of a black hole is also associated with its horizon radius. 
One way to calculate the black-hole entropy is through Wald's formula \cite{Wald:1993nt, Iyer:1994ys}, which links the Noether charge on the horizon with the entropy of the black hole. 
This is expressed as
\begin{equation}
S=-2\pi\oint d^2 x \sqrt{h_2}\left( \frac{\partial \mathcal{L}}{\partial R_{abcd}} \right)_{\mathcal{H}}\hat{\epsilon}_{ab}\,\hat{\epsilon}_{cd}\, ,
\end{equation}
where $\hat{\epsilon}_{ab}$ represents the binormal to the horizon's surface $\mathcal{H}$, $h_2$ denotes the determinant of the 2-dimensional projected metric on $\mathcal{H}$, and $\mathcal{L}$ stands for the Lagrangian density of the theory. 
Using this equation, it can be shown that in our theory, the entropy follows the Bekenstein-Hawking formula $S=\mathcal{A}/4$. Thus,
\begin{equation}\label{eq: entr-ratio}
\frac{S}{S_{\rm Sch}}=\left( \frac{r_h}{2M}\right)^2,
\end{equation}
where $S_{\rm Sch}$ is the entropy of the corresponding Schwarzschild black hole with mass $M$. 
Pertaining now to the black-hole solutions with regular hair, meaning their hair originates from a regular scalar field, it is easy to see from Fig. \myref{fig: rh-plots}{subf: real-rh} and eq. \eqref{eq: entr-ratio} that the ratio $S/S_{\rm Sch}$ declines as the value of $q/r_h$ rises. 
Consequently, as these solutions acquire more hair, they become less thermodynamically stable.
Conversely, one can infer from Fig. \myref{fig: rh-plots}{subf: phant-rh} that black holes with phantom hair become more thermodynamically stable as they gain hair.
Hence, from a thermodynamic perspective, ultra-sparse black holes appear much more stable than ultra-compact ones. 

\begin{figure}[t]
    \centering
    \includegraphics[width=0.6\textwidth]{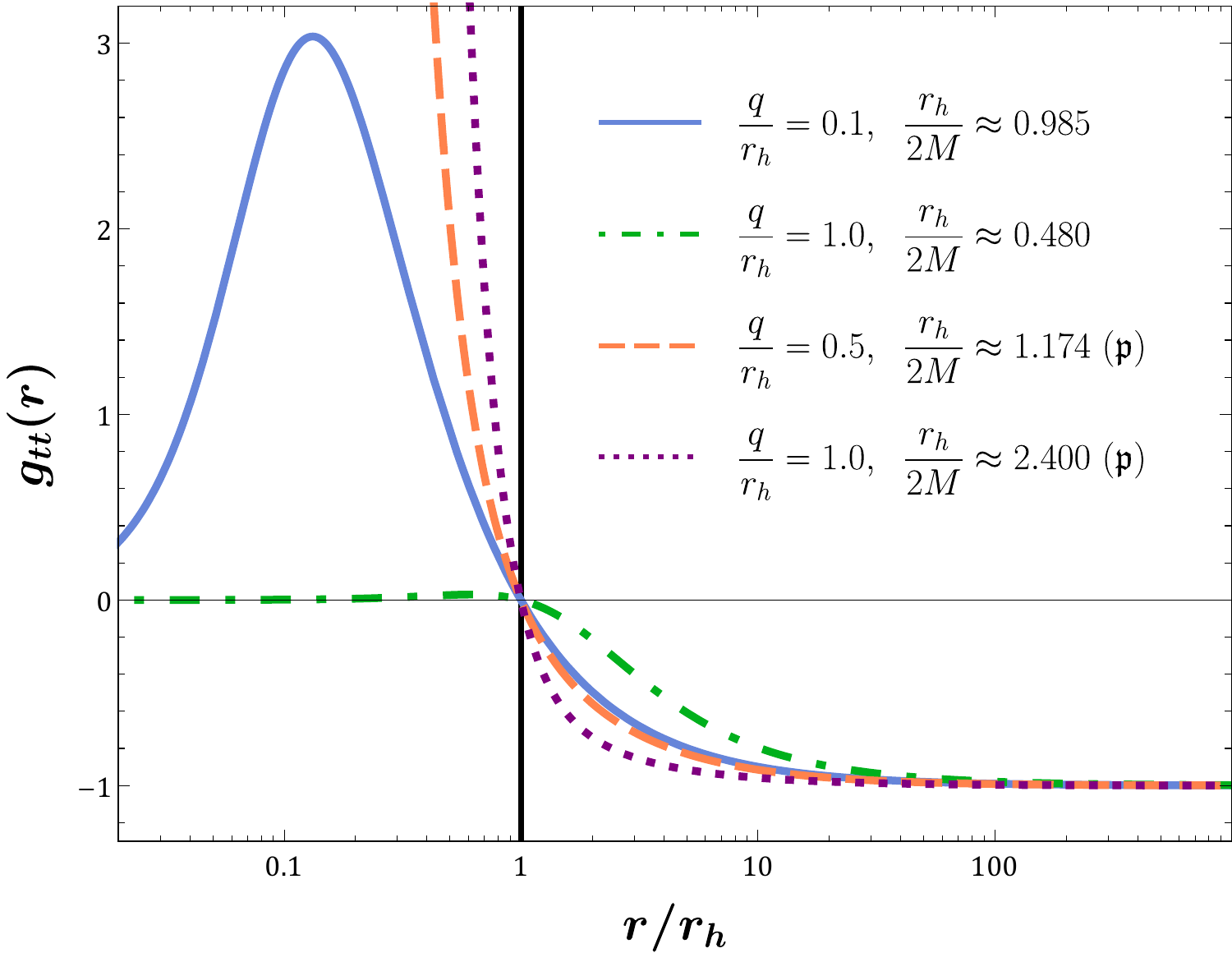}
    \caption{Graphs of the function $g_{tt}(r)$ in terms of $r/r_h$ for both normal and phantom solutions and for various values of the parameter $q/r_h$. 
    The graphs containing the symbol $\mathfrak{p}$ in their legend, refer to phantom solutions.
    The horizontal axis is logarithmic.}
    \label{fig: hor-plot}
\end{figure}

Let us now turn our attention back to the analysis of the causal structure of the spacetime. 
Fig. \ref{fig: hor-plot} displays the function $g_{tt}(r)$ plotted against the dimensionless radial quantity $r/r_h$ for different values of the parameter $q/r_h$. 
We have selected $q/r_h = 0.1,\, 1.0$ for the normal solutions, and $q/r_h=0.5,\, 1.0$ for the phantom ones. 
By utilizing eqs. \eqref{eq: rh-real} and \eqref{eq: rh-phant}, we can readily compute that these parameter values result in $r_h/(2M) \approx 0.985,\, 0.480$ when the scalar field is regular, and $r_h/(2M) \approx 1.174,\, 2.400$ when the scalar field is phantom. 
Note also, that by substituting the black-hole mass $M$ from eqs. \eqref{eq: rh-real} and \eqref{eq: rh-phant} into the expressions \eqref{eq: B-real} and \eqref{eq: B-phant}, respectively, we have kept the same horizon radius in all cases.
So, the first thing that one notices in Fig. \ref{fig: hor-plot} is that beyond the black-hole horizon, the function $g_{tt}(r)$ converges rapidly to its asymptotic value, regardless of whether the field is regular or not, since $g_{tt}(100\, r_h)\approx -1$.
However, the cases of regular and phantom scalar fields exhibit a significant difference in their behavior; the function $g_{tt}(r)$ of the phantom solutions manifests a singular behavior at $r=0$, while for the normal solutions, $g_{tt}(r\rightarrow 0)$ remains finite.
Finally, it is apparent from the graphs in Fig. \ref{fig: hor-plot}, that an observer would only be able to gravitationally distinguish between phantom and normal solutions if they were in close proximity to the horizon.

\begin{figure}[t]
    \centering
    \begin{subfigure}[b]{0.48\textwidth}
    \includegraphics[width=1\textwidth]{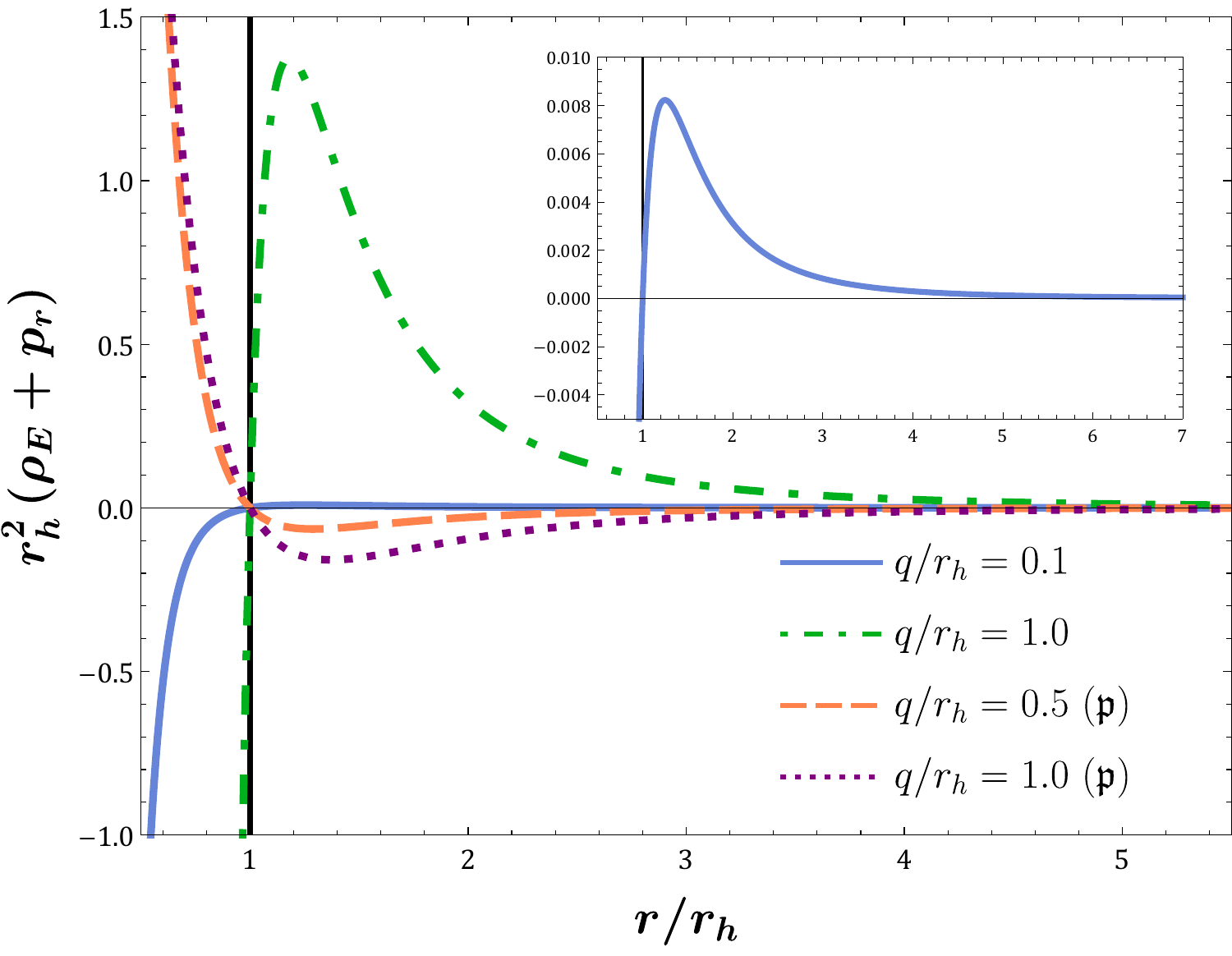}
    \caption{\hspace*{-3em}}
    \label{subf: NEC}
    \end{subfigure}
    \hfill
    \begin{subfigure}[b]{0.49\textwidth}
    \includegraphics[width=1\textwidth]{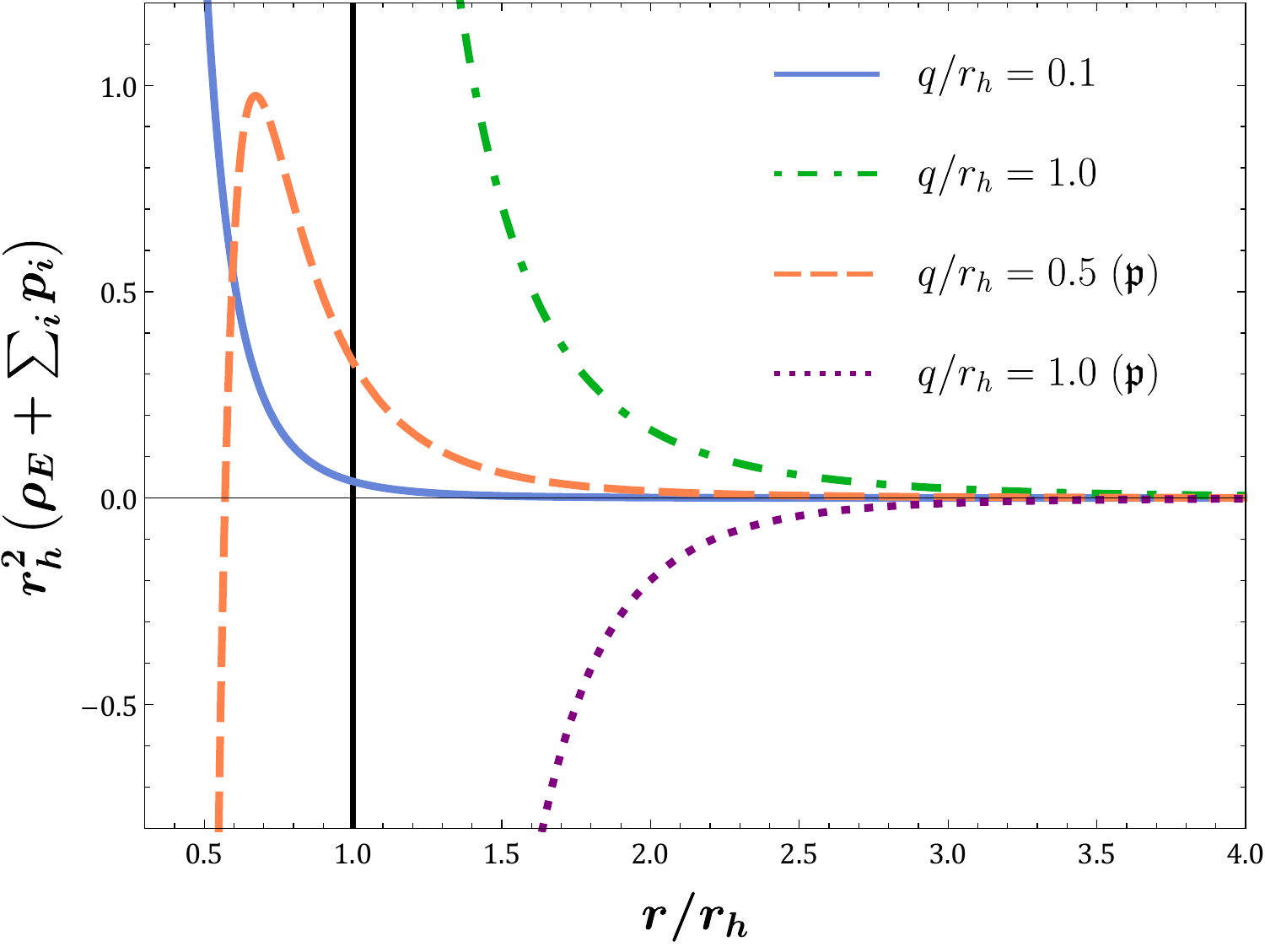}
    \caption{\hspace*{-2.8em}}
    \label{subf: SEC}
    \end{subfigure}
    \caption{Graphs of the quantities (a) $r_h^2(\rho_E+p_r)$ and (b) $r_h^2\big( \rho_E +\sum_i p_i\big)$ in terms of $r/r_h$, for both normal and phantom solutions and for various values of the parameter $q/r_h$. The graphs containing the symbol $\mathfrak{p}$ in their legend, refer to phantom solutions. }
    \label{fig: EC}
\end{figure}

Having examined the geometrical characteristics of the line-element \eqref{metr-ans}, it is now crucial to investigate the energy conditions satisfied by the stress-energy tensor \eqref{stress-ten} associated with the scalar field. 
The stress-energy tensor $T^{(\Phi)\mu}{}_\nu$, in its entirety, is described by three physical quantities: the energy density $\rho_E=-T^{(\Phi)t}{}_t$, the radial pressure $p_r=T^{(\Phi)r}{}_r$, and the tangential pressure $p_\theta=T^{(\Phi)\theta}{}_\theta=T^{(\Phi)\varphi}{}_\varphi$. 
By noticing eq. \eqref{stress-ten}, one can readily deduce that $p_\theta=-\rho_E$, whereas $p_r=w_r(r) \rho_E$, with the function $w_r(r)$ defining the equation of state of the radial pressure $p_r$. 
In Figs. \myref{fig: EC}{subf: NEC} and \myref{fig: EC}{subf: SEC}, we display the graphs of $r_h^2 (\rho_E+p_r)$, and $r_h^2\big(\rho_E+\sum_i p_i\big) \equiv r_h^2(\rho_E + p_r + 2p_\theta) = r_h^2(p_r + p_\theta)$, respectively, plotted against the dimensionless radial quantity $r/r_h$. 
In all graphs, we have multiplied the depicted functions with $r_h^2$ to make the resulting quantities dimensionless and scale invariant as well.
Note that the aforementioned graphs are demonstrated for the same values of the dimensionless parameter $q/r_h$ as in Fig. \ref{fig: hor-plot}.
In addition to the above, in all formulas, the mass $M$ has been replaced using eq. \eqref{eq: rh-real}, for the normal, and eq. \eqref{eq: rh-phant}, for the phantom solution, in order to retain the same horizon radius in all cases.

It is evident from Fig. \myref{fig: EC}{subf: NEC} that in both normal and phantom cases, the quantity $\rho_E+p_r$ nullifies at $r=r_h$ independently of the value of $q/r_h$.
Although this behavior can indeed be verified analytically as well, we are not going to elaborate more on this since we wish to keep the discussion as concise as possible.
By utilizing eq. \eqref{stress-ten}, one may confirm by their own the preceding assertion.
From Figs. \myref{fig: EC}{subf: NEC} and \myref{fig: EC}{subf: SEC}, we clearly observe that a regular scalar field always satisfies the null and strong energy conditions in the causal region of spacetime, that is outside the black-hole horizon $r_h$.
This is due to the fact that for $r>r_h$, both conditions $\rho_E+p_i\geq 0$ and $\rho_E+\sum_i p_i\geq 0$\,\footnote{For a perfect fluid $\rho+\sum_i p_i=\rho+3p$, and the second condition would have been $\rho+3p\geq 0$.} ($p_i=\{p_r,\, p_\theta\}$) are satisfied everywhere and independently of the value of $q/r_h$.
Conversely, a phantom scalar field violates both null and strong energy conditions.
Although this violation is more often met in wormhole solutions \cite{Morris:1988cz,Visser:1995cc,Kanti:2011jz,Kanti:2011yv,Antoniou:2019awm}, it is also anticipated from black-hole solutions with a phantom scalar hair.
The reason is that by definition, the kinetic term of phantom scalar fields comes with the opposite sign, as it was discussed in \eqref{eq: phan-act}.
This sign alone is the one that ultimately leads the phantom fields to violate the null energy condition, and by its turn the strong energy conditions as well. Finally, it is important to mention that both normal and phantom cases violate either one or both weak energy conditions. This is indeed anticipated in order for the solutions to evade the no-scalar-hair theorem. This is also the case even if we consider the complete Einstein-scalar-Gauss-Bonnet theory \eqref{actiongb} or any other theory that evades the no-scalar-hair theorem.

We will now shift our attention to the \textit{no-scalar hair theorems} \cite{NH-scalar1,NH-scalar2,Bekenstein}, which prohibit the existence of black holes in the presence of a scalar field.
As it was discussed in our previous work \cite{Bakopoulos:2021dry}, for the field theory considered here, the no-scalar hair theorem \cite{Bekenstein} is solely dependent on the correlation between the sign of the potential $V(\Phi)$ and the sign of the kinetic term of the scalar field.
By employing the scalar equation \eqref{sc-eq}, as discussed in \cite{Bakopoulos:2021dry}, one obtains the conditions
\eq$\label{eq: nh-real}
\int_{r_h}^\infty dx^4 \sqrt{-g}\, \pa_\Phi V\left( \partial_\mu \Phi\,\partial^\mu \Phi+ V \right)=0\,,$
and
\eq$\label{eq: nh-phant}
\int_{r_h}^\infty dx^4 \sqrt{-g}\, \pa_{\tilde{\Phi}} V_{\mathfrak{p}}\left(-\partial_\mu \tilde{\Phi}\,\partial^\mu \tilde{\Phi}+ V_\mathfrak{p} \right)=0\,,$
for a regular and a phantom scalar field, respectively.
In the above relations, the terms of the form $\left[\sqrt{-g}\,V\partial^\mu\Phi \right]_{r_h}^\infty$ vanish in both boundaries. 
On the black-hole horizon, they vanish since the factor $B(r)$ that appears through the derivative of the scalar field with respect to $r$,\,\footnote{$\partial^\mu\Phi=\del^{\mu r}g^{rr}\pa_r\Phi=\del^{\mu r}B(r)\pa_r\Phi$ and $B(r_h)=0$ by definition.}  whilst at infinity the scalar potentials vanish due to the fact that the black holes considered here are asymptotically flat.
Notice that in the case of a regular scalar field, the first term in \eqref{eq: nh-real} can be written as $\pa_\mu\Phi\, \pa^\mu\Phi = g^{rr}\Phi'^2 >0$, while in the phantom case, the corresponding term reads $-\pa_\mu \tilde{\Phi}\, \pa^\mu \tilde{\Phi}= - g^{rr}\tilde{\Phi}'^2<0$.
Consequently, eq. \eqref{eq: nh-real} can be satisfied only if the scalar potential is negative-definite, that is $V(\Phi)<0$, whereas eq. \eqref{eq: nh-phant} indicates that the potential of the phantom scalar field should be positive, i.e. $V(\tilde{\Phi})>0$.

\begin{figure}[t]
    \centering
    \includegraphics[width=0.6\textwidth]{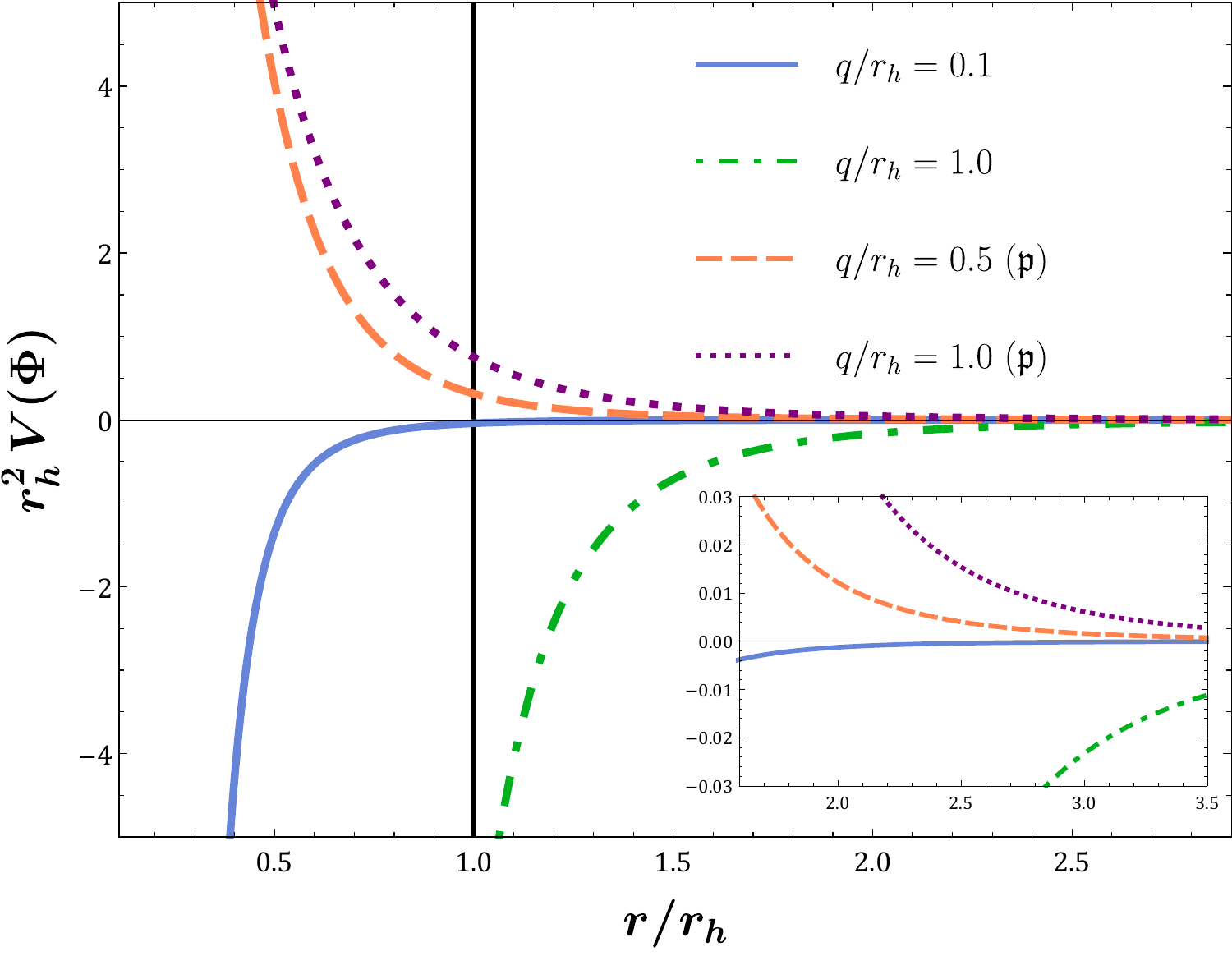}
    \caption{The potential $r_h^2 V(\Phi(r))$ in terms of $r/r_h$ for different values of the parameter $q/rh$.  
    The graphs containing the symbol $\mathfrak{p}$ in their legend, refer to phantom solutions. }
    \label{fig: V-plot}
\end{figure}

In Fig. \ref{fig: V-plot}, we display the graphs of the potentials $V(\Phi)$ and $V(\tilde{\Phi})$ in terms of the dimensionless radial quantity $r/r_h$.
Similarly to what we did in the previous figure, we have multiplied the potential with $r_h^2$ to obtain a dimensionless and scale invariant quantity.
In accordance with what we discussed previously, it is clear that the potential regarding a regular scalar field is consistently negative, whereas, for a phantom scalar field, it is consistently positive.
Finally, we see that both cases lead to asymptotically flat solutions since both potentials converge rapidly to zero, which is their asymptotic value.



\section{Slow-rotating black holes}
\label{Sec: rot-BHs}

The objective of this section is to extend the static line-element \eqref{metr-ans} to encompass slowly rotating black-hole solutions. This will be accomplished by treating the rotating solutions as an axisymmetric perturbation on the static and spherically symmetric spacetime \eqref{metr-ans}. As background solutions, we will use the two solutions that we derive in the previous sections. The method we follow was first proposed by Hartle \cite{Hartle:1967he} in the framework of General Relativity and generalized by Pani and Cardoso for scalar-tensor theories \cite{Pani:2009wy}. 
Thus, we are led to consider the line-element
\eq$\label{slow-metr}
{ds}^2=-e^{A(r)}B(r)\,{dt}^2+\frac{dr^2}{B(r)}+r^2\left\{{d\theta}^2+\sin^2\theta\left[d\varphi-\varepsilon\, \omega(r)\, dt\right]^2\right\}.$
The above expression involves a dimensionless auxiliary parameter $\varepsilon$ that enables us to regulate the perturbations, while
the function $\omega(r)$ is directly related to the angular velocity that a locally stationary observer would obtain at distance $r$, due to the rotational frame-dragging.
In general, one may allow the $omega$ function to have an angular $\theta$ dependence or even assume a scalar field perturbation of the form $\Phi^{tot}(r,\theta)=\Phi(r)+\varepsilon\, \Phi_1(r,\theta)$. However, in a previous work of ours \cite{Bakopoulos:2021dry}, we have shown that at first order in $\varepsilon$, the angular velocity depends only on the radial coordinate while the scalar perturbation vanishes. By evaluating the expansion as $\varepsilon$ approaches zero, we discover that the unaltered set of equations admits the general solution \eqref{eq: B-r} described in the previous section. At first order in $\varepsilon$ though, i.e. $\mathcal{O}(\varepsilon)$ in the expansion, the only independent field equation comes from the $(t,\varphi)$ component and it can be presented in the following manner:
\gat$
\label{pert-r}
\left[r^4 e^{-\frac{A(r)}{2}}\omega'(r)\right]'=0\,.$
Therefore, it is straightforward to derive that
\begin{equation}\label{pert-r1}
    \omega'(r)=\frac{\omega_1\, e^{\frac{A(r)}{2}}}{r^4}\,,
\end{equation}
where $\omega_1$ is an integration constant.

We observe that, in our theory, the angular velocity $\omega(r)$ is solely determined by the metric function $A(r)$, and is independent of whether we are referring to black-hole solutions emanating from a regular or a phantom scalar field. 
By demanding the asymptotic expansion of the angular velocity $\omega(r)$ to be identical to the angular velocity of the slowly rotating Schwarzschild solution, namely $\omega_{\rm Sch}(r)=2 J/r^3$, we readily find that $\omega_1=-6J$, with $J$ being the angular momentum of the black hole \cite{Pani:2009wy}.
The integration of eq. \eqref{pert-r1} leads to the expression
\begin{equation}\label{omegaint}
   \omega(r)=\omega_0- 6J\int_0^r \frac{1}{x^4}\,  e^{\frac{A(x)}{2}} dx\,. 
\end{equation}
In the above, the integration constant $\omega_0$, is merely specified by imposing the condition that the angular velocity becomes zero as $r$ approaches infinity.
Thus, we obtain
\begin{equation}
    \omega_0=6J\int_0^\infty \frac{1}{r^4}\,  e^{\frac{A(r)}{2}} dr\,. 
\end{equation}
By substituting the metric function $A(r)$, the integral (\ref{omegaint}) may be expressed in terms of the hypergeometric function, that is 
\begin{equation}
\label{eq: ang-vel}
\omega(r)=\omega_0  -  \frac{6 J r^{\xi -3}  }{q^\xi(\xi -3)}\,_2F_1\left(\frac{\xi -3}{2},\frac{\xi }{2};\frac{\xi
   -1}{2};-\frac{r^2}{q^2}\right)\,.
\end{equation}

Our attention will now be directed towards the solutions that were studied throughout the present manuscript, and specifically the solution for a regular scalar field with $\xi=5$, as well as the solution related to a phantom scalar field characterized by $\xi=-2$.
For $\xi=5$, eq. \eqref{eq: ang-vel} leads to the simple relation
\eq$\label{eq: ang-vel-real}
\omega(r)=\frac{2J}{r^3}\left(1+\frac{q^2}{r^2}\right)^{-3/2}\,,
$
while for $\xi=-2$ one finds that
\eq$\label{eq: ang-vel-phant}
\omega_\mathfrak{p}(r)=\frac{2J}{r^3}\left(1+\frac{3q^2}{5\,r^2}\right)\,.$
%
%
We shall now examine the relative angular velocities of a slowly rotating black hole with either regular or phantom hair, and a slowly rotating Schwarzschild black hole of equivalent mass. 
To this end, we will determine the ratio of their respective angular velocities at the horizon for each black hole. 
This calculation can be performed utilizing equations \eqref{eq: ang-vel-real} and \eqref{eq: ang-vel-phant} along with equations \eqref{eq: rh-real} and \eqref{eq: rh-phant}. 
Below we present the resulting expressions:
\eq$\label{eq: rat-ang-real}
\frac{\omega(r_h)}{\omega_{\rm Sch}(2M)}=\left( 1 + 3\,\frac{q^2}{r_h^2} + \frac{17}{9}\frac{q^4}{r_h^4} \right)^{3} \left( 1 + \frac{q^2}{r_h^2} \right)^{-6}\,,$
and
\eq$\label{eq: rat-ang-phant}
\frac{\omega_\mathfrak{p}(r_h)}{\omega_{\rm Sch}(2M)}= \left( 1 - \frac{1}{3}\frac{q^4}{r_h^4} \right)^3 \left( 1 + \frac{3}{5}\frac{q^2}{r_h^2}\right)^{-2}\,.$
Here, $\omega(r_h)/\omega_{\rm Sch}(2M)$ denotes the ratio regarding normal black holes, while $\omega_\mathfrak{p}(r_h)/\omega_{\rm Sch}(2M)$ corresponds to the ratio of phantom black holes.

\begin{figure}[t]
    \centering
    \begin{subfigure}[b]{0.49\textwidth}
    \includegraphics[width=1\textwidth]{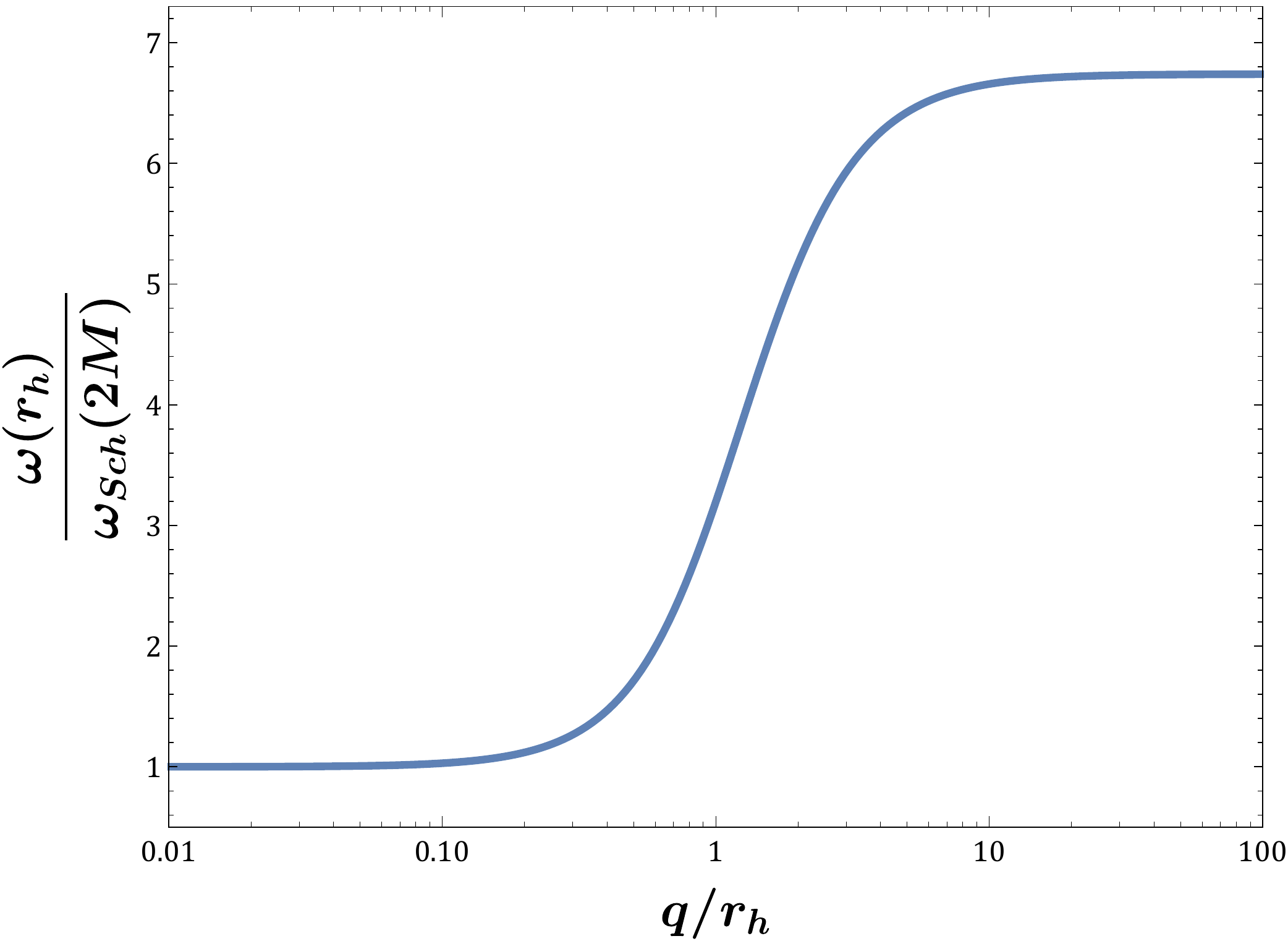}
    \caption{\hspace*{-3em}}
    \label{subf: omeg-real}
    \end{subfigure}
    \hfill
    \begin{subfigure}[b]{0.49\textwidth}
    \includegraphics[width=1\textwidth]{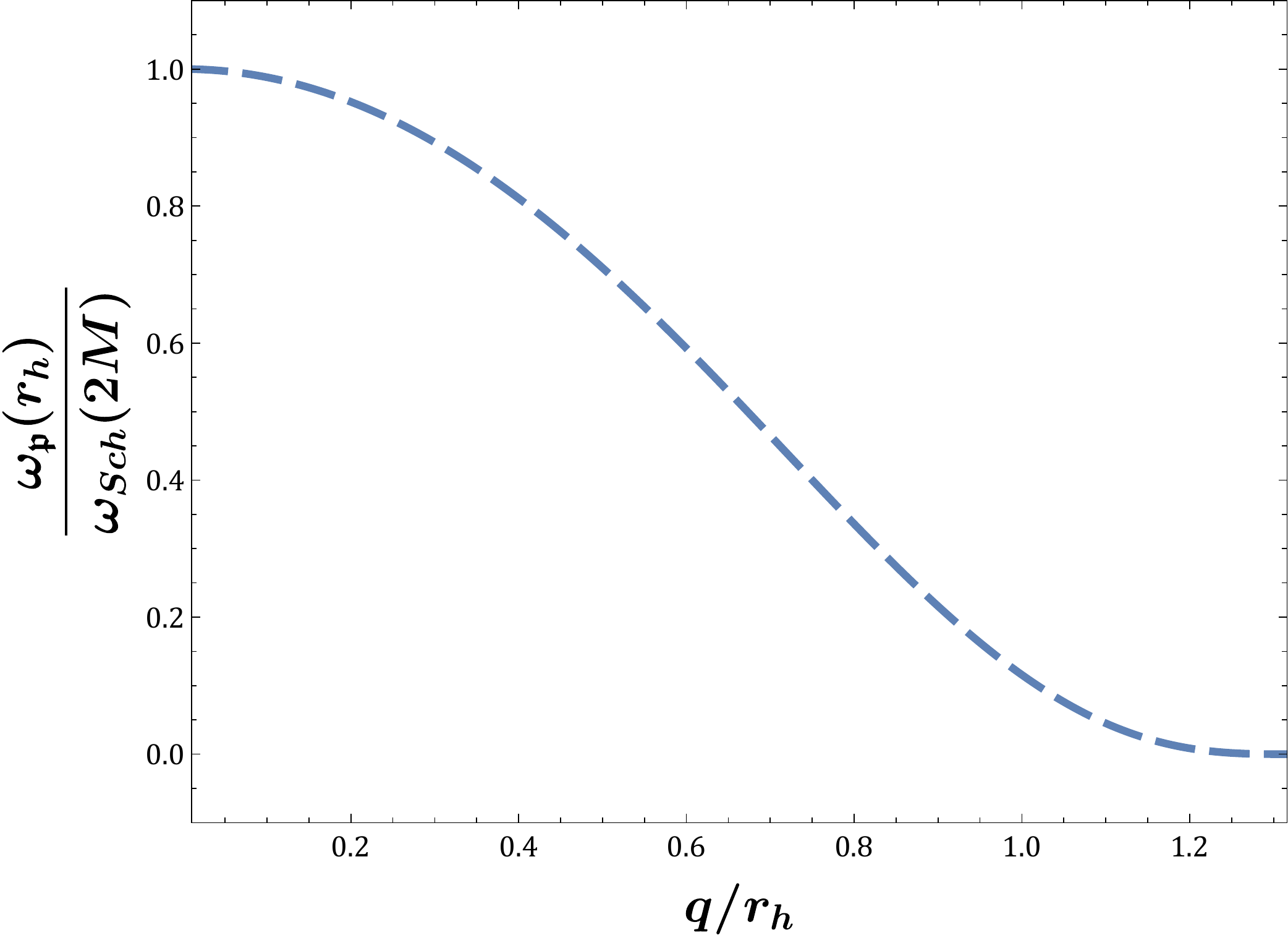}
    \caption{\hspace*{-4em}}
    \label{subf: omeg-phant}
    \end{subfigure}
    \caption{Graphs of the ratio $\omega(r_h)/\omega_{Sch}(2M)$ in terms of $q/r_h$ for (a) the normal and (b) the phantom solutions.
    The horizontal axis in (a) is logarithmic.}
    \label{fig: omeg}
\end{figure}

Figs. \myref{fig: omeg}{subf: omeg-real} and \myref{fig: omeg}{subf: omeg-phant} display the graphs of eqs. \eqref{eq: rat-ang-real} and \eqref{eq: rat-ang-phant}, respectively, plotted against the dimensionless quantity $q/r_h$. 
We consider the case in which the two compact objects are of identical mass $M$. 
As previously shown in Fig. \ref{fig: rh-plots}, the values of their horizon radii can differ significantly depending on the parameter $q/r_h$. 
In both cases, we observe that the angular velocities are equal to $\omega_{\rm Sch}(2M)$ for small values of $q/r_h$. 
This result is expected since for small values of $q/r_h$, the horizon radius of both black-hole solutions is identical to the horizon radius of the corresponding Schwarzschild black hole with the same mass, i.e. $r_h=2M$ (refer to Fig. \ref{fig: rh-plots}). 
It is also apparent from equations \eqref{eq: ang-vel-real} and \eqref{eq: ang-vel-phant} that in the limit $q\rightarrow 0$, the leading term in both cases is the same as that of the slowly rotating Schwarzschild black hole, namely $2J/r^3$.

To be more specific, Fig. \myref{fig: omeg}{subf: omeg-real} reveals that the ratio $\omega(r_h)/\omega_{\rm Sch}(2M)$ remains almost equal to unity for  $q/r_h\lessapprox 0.2$. 
From $q/r_h\approx 0.2$ to $q/r_h\approx 7$, the angular velocity $\omega(r_h)$ grows seven times compared to $\omega_{\rm Sch}(2M)$, while for the values $q/r_h\gtrapprox 7$, the relative value of the angular velocities remains essentially constant. 
A similar behavior was also observed in \cite{Bakopoulos:2021dry} for hairy black-hole solutions, with their scalar hair being of a coulombic form.
On the contrary, as shown in Fig. \myref{fig: omeg}{subf: omeg-phant}, the angular velocity of a given phantom black hole is consistently smaller than the corresponding angular velocity of a Schwarzschild black hole with the same mass. 
Additionally, the same graph illustrates that as the value of $q/r_h$ increases, the angular momentum of the phantom black holes decreases until it reaches zero at $q/r_h=\sqrt[4]{3}$. 
It should be noted that the phantom solutions considered here are ultra-sparse, and thus, as $q/r_h$ increases, the horizon radius also increases until it approaches infinity at $q/r_h=\sqrt[4]{3}$.
The aforementioned behavior of the angular velocity, of either normal or phantom black-hole solutions, can be naively understood, even by using a reasoning based on Classical Mechanics.
It is common knowledge that an increase in the horizon radius of a black hole, with the mass kept constant, corresponds to an increase in its moment of inertia, which in its turn makes the rotation of such an object harder.
Conversely, for an object with a smaller horizon radius, its moment of inertia would be smaller, and its rotation would be easier.


\section{Stability}
\label{Sec: stab}

In this section, we will analyze the linear stability of the black-hole solution described by eqs. (\ref{eq: A-r}, \ref{eq: phi-r}, \ref{eq: B-r}).
To this end, we consider small perturbations $h_{\mu\nu}$ in the background spacetime $g_{\mu\nu}$, namely $|h_{\mu\nu}|\ll |g_{\mu\nu}|$. 
The combined effect of the background metric $g_{\mu\nu}$ and spacetime perturbations $h_{\mu\nu}$ will henceforth be referred to as the total metric tensor $g^{tot}_{\mu\nu}$ defined as
\begin{equation}
    \label{metr-tot}
g^{tot}_{\mu\nu}=g_{\mu\nu}+h_{\mu\nu}\,.
\end{equation}
The approach we will employ to study the stability is the same as the one initially introduced by Regge and Wheeler in 1957 \cite{Regge:1957td}, which was later refined and extended by Zerilli and Vishveshwara in 1970 \cite{Zerilli:1970se, Vishveshwara:1970cc, Zerilli:1971wd}.
Given, however, that the Lagrangian density of our theory also contains the kinetic and potential terms of a scalar field, it becomes necessary to consider scalar-field perturbations as follows:
\eq$\label{Phi-tot}
\Phi^{tot}=\Phi+\del\Phi\,.$

The perturbations are classified into two distinct categories based on their parity properties: perturbations exhibiting \textit{odd} parity $(-1)^{L+1}$, which are commonly referred to as axial perturbations, and perturbations exhibiting \textit{even} parity $(-1)^L$, which are typically referred to as polar perturbations. 
Here, $L$ represents the angular momentum associated with the specific perturbation mode. 
For simplicity, in the context of this work, we will restrict our analysis to the odd sector. 
Due to their complexity, the even perturbations will be examined in a future work. 
Although the line-element \eqref{metr-ans} describes static black-hole solutions, in principle, the spacetime perturbations $h_{\mu\nu}$ depend on all spacetime coordinates. 
Using the method of separation of variables, we may express the odd perturbations in the Regge-Wheeler gauge \cite{Regge:1957td} as 
\eq$\label{P-odd}
h^{\rm odd}_{\mu \nu}= \left[
\begin{array}{cccc} 
0 & 0 &0 & h_0(r) 
\\ 0 & 0 &0 & h_1(r)
\\ 0 & 0 &0 & 0
\\ h_0(r) & h_1(r) &0 &0
\end{array}\right] e^{-i k t}
\sin\theta\,\partial_\theta P_L(\cos\theta)\,,\hspace{2em}\del\Phi=0\,.$
The decomposition into modes of fixed energy is achieved using the term $\exp(-ikt)$, where $k$ denotes the frequency of the mode. 
Similarly, the decomposition into modes with fixed angular momentum $L$ is attained using the Legendre polynomials $P_L(\cos\theta)$. 
Note also that the perturbations of \eqref{P-odd} are expressed in the canonical gauge as it is often called.
For more details regarding spacetime perturbations, one is referred to \cite{Regge:1957td,Zerilli:1970se, Vishveshwara:1970cc, Zerilli:1971wd}.
The radial part of the background metric perturbations are encoded inside the functions $h_0(r)$ and $h_1(r)$.

After the substitution of eqs. (\ref{metr-tot}-\ref{P-odd}) into the field equations and by retaining only the linear terms in $h_{\mu\nu}$, one finds that for the odd metric perturbations, there merely exist two independent equations, $(\theta,\varphi)$ and $(r,\varphi)$.
The $(\theta,\varphi)$ equation is evaluated to be
\begin{equation}
    \left\{\frac{e^A}{4r^2} \left[h_1(A'B+2B')+2B\,h_1')\right]-\frac{ik}{2r^2B}\, h_0 \right\} \left[2 \cot \theta\, \pa_\theta P_L(\cos\theta)+L  (L +1) P_L(\cos\theta)\right]=0\,,\label{perteq1}
\end{equation}
while the $(r,\varphi)$ is of the form
\begin{align}
    &\bigg\{ h_1 \bigg[\frac{e^A}{4r^4} \left(3 r^2 A' B'+2 r^2 B''+2 L  (L +1)\right)+\frac{e^A B}{2r^4} \left(r^2 A''+\frac{r^2 A'^2}{2}- r A'+ r^2 \Phi'^2-2\right)- \frac{k^2}{r^2B}\bigg]\nonumber\\[2mm]
   &\hspace{0.5em}- \frac{i k}{2r^2B}\, h'_0+\frac{ ik}{r^3B}\, h_0  \bigg\}\pa_\theta P_L(\cos\theta)=0.\label{perteq2}
\end{align}
It should be mentioned at this point that for $L=0$, $P_0(\cos\theta)=1$, and both eqs. \eqref{perteq1} and \eqref{perteq2} become identically zero. 
Consequently, there are no odd-parity perturbations for $L=0$. 
Similarly, in the case of $L=1$, where $P_1(\cos\theta)=\cos\theta$, we observe that eq. \eqref{perteq1} is satisfied identically as well. 
And although it may seem that we have \eqref{perteq2} to work with, one can readily demonstrate that eq. \eqref{perteq2} can be completely gauged away by first performing a coordinate transformation of the form
\eq$\label{new-gauge}
x'^\mu=x^\mu+i\, \del^{\mu}{}_\varphi\, \frac{ e^{-ikt}}{k\,r^2}\,h_0(r)\,,$
and then by redefining the function $h_1(r)$ via the relation
\eq$\label{h1-redef}
h_1(r)=i\,\frac{rh_0'(r)-2h_0(r)}{k\,r}\,.$
Hence, odd-parity modes are only present for $L\geq 2$.  

As the angular component of the aforementioned equations is either non-vanishing or non-singular for $L\geq 2$, we may concentrate solely on their radial part. 
By algebraically solving eq. (\ref{perteq1}) with respect to $h_0(r)$ and substituting the obtained result into eq. (\ref{perteq2}), one obtains a second-order differential equation for $h_1(r)$. 
If we now define a new radial function $\Psi(r)$ via
\begin{equation}
    h_1(r)=\dfrac{r\, \Psi(r)}{B(r)\, e^{A(r)/2}}\,,
\end{equation}
and also impose the tortoise coordinate $r^*$ with the use of the transformation $dr^*=dr\,e^{-A(r)/2}/B(r)$, the aforementioned differential equation takes a Schr\"{o}dinger-like form, namely
\begin{equation}
\label{Schr-eq}
    \frac{d^2\Psi(r^*)}{dr^{*2}}+\left[ k^2 - \mathcal{V}(r)   \right]\,\Psi(r^*)=0\,.
\end{equation}
In the above relation, the potential is determined by
\begin{equation}
    \mathcal{V}(r)=\frac{e^A B}{2r} \left\{B' \left(3 r A'-2\right)+B \left[2 r \left(A''+\Phi'^{\,2}\right)+r
   A'^2-3 A'\right]+2 r B''+\frac{2 L  (L +1)}{r}\right\}\,.\label{potsch}
\end{equation}
For more information about the derivation of the above equation see \cite{Bakopoulos:2021dry}.
The tortoise coordinate facilitates a coordinate transformation of the region $[r_h,+\infty)$, mapping it to the interval $(-\infty,+\infty)$.
In this way, divergences which were previously appeared on the black-hole horizon $r_h$, are now moved to an infinite distance, and specifically to minus infinity. 
This transformation enables the tortoise coordinate to serve as a parameterization for the entire exterior spacetime of the black hole.

\begin{figure}[t]
    \centering
    \begin{subfigure}[b]{0.48\textwidth}
    \includegraphics[width=1\textwidth]{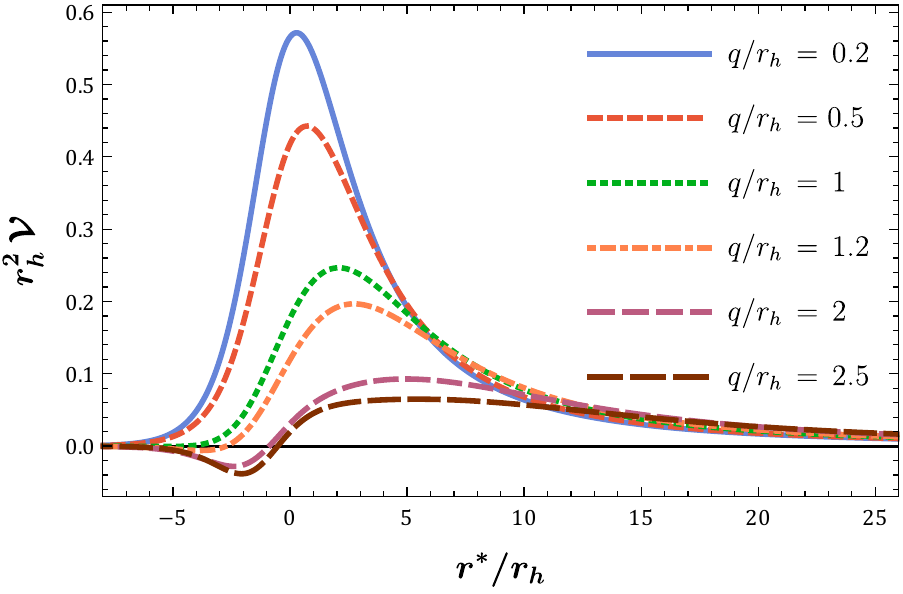}
    \caption{\hspace*{-3em}}
    \label{subf: stab-real}
    \end{subfigure}
    \hfill
    \begin{subfigure}[b]{0.49\textwidth}
    \includegraphics[width=1\textwidth]{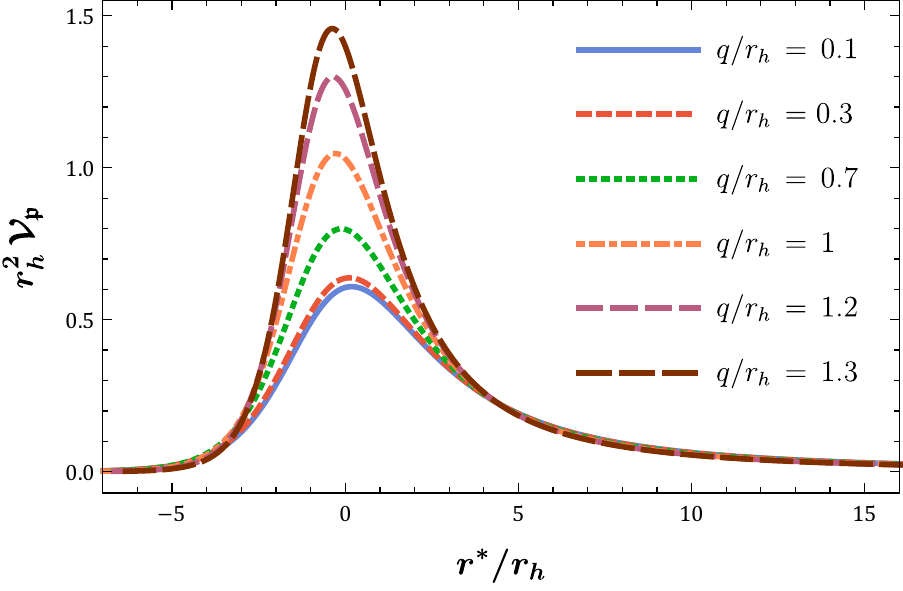}
    \caption{\hspace*{-2.8em}}
    \label{subf: stab-phant}
    \end{subfigure}
    \caption{The potential $r_h^2\, \mathcal{V}$ in terms of the tortoise coordinate $r^*/r_h$ for a family of (a) normal and (b) phantom solutions.
    The value of the angular momentum $L$ is equal to 2, while the dimensionless parameter $q/r_h$ varies. }
    \label{fig: stab}
\end{figure}

Since our spotlight is on the stability of the black-hole solutions, there is no need to solve eq. (\ref{Schr-eq}). 
The time evolution factor $\exp(-i k t)$ simplifies our task, requiring us only to determine if the frequency $k$ is purely imaginary. 
A negative eigenvalue $k^2<0$ indicates an unstable mode that corresponds to a bound state of the Schr\"{o}dinger equation \eqref{Schr-eq}. 
Thus, in this case the frequency $k$ is purely imaginary, and the mode experiences exponential growth due to the presence of the term $\exp(-ikt)$.
According to \cite{Buell}, if the potential $\mathcal{V}(r^*)$ approaches zero as $r^*\rightarrow\pm\infty$, then the requirement to exist at least one bound state is
\begin{equation}
    \int_{-\infty}^{+\infty}\mathcal{V}(r^*)\,dr^*<0\,.
\end{equation}
However, it has been argued by the authors of \cite{Buell} that the existence of a bound state is not ruled out even if the integral mentioned above is positive. 
This can be understood intuitively as well, by simply imagining that for potentials with shapes similar to those shown in Fig. \myref{fig: stab}{subf: stab-real} for $q/r_h\geq 1.2$, there is no obstacle for a bound state to exist in the region where the potential is negative-definite and forms a well. 
Hence, we assert that any solution with a negative region in its $\mathcal{V}(r^*)$ potential must contain at least one unstable mode. 
Notice also, that due to the fact that the term $L(L+1)$ in eq. (\ref{potsch}) introduces a positive angular barrier to the potential, it becomes sufficient to study the stability of our black-hole solutions in the mode where the value of the angular momentum is minimum, i.e., $L=2$. 
In Fig. \myref{fig: stab}{subf: stab-real} we depict the graphs of the potential $\mathcal{V}(r^*)$ in terms of the dimensionless quantity $r^*/r_h$, for a family of six normal solutions. 
Respectively, Fig. \myref{fig: stab}{subf: stab-phant} illustrates the graphs of the potential for a family of six phantom solutions. 
By purely observing the form of the graphs, one can promptly deduce that normal solutions with $\xi=5$ are stable for $q/r_h \lesssim 1$, 
while the phantom solutions with $\xi=-2$ proves to be stable $\forall\, \frac{q}{r_h}\in(0,\sqrt[4]{3})$. 
This outcome is consistent with the thermodynamic analysis presented in Sec. \ref{Sec: theory}. Specifically, it is observed that ultra-sparse black holes exhibit greater stability relative to ultra-compact ones, both under odd spacetime perturbations and from a thermodynamic perspective.



\section{Conclusions}
\label{Sec: concl}

In this work, we have investigated a basic theory featuring a scalar field that is minimally coupled to gravity. 
The Lagrangian density of the scalar field contains both its kinetic and potential terms.
Having adopted a spherically symmetric form for the metric tensor and a specific expression for one of its functions, we were then able to solve the field equations and determine explicitly the expressions of the unknown functions, including the scalar field and its potential.
It was proven that depending on the type of the scalar field, regular or phantom, the resulting solutions may describe either ultra-compact or ultra-sparse black holes.
We then generalized these solutions into slowly rotating ones by utilizing the method developed in \cite{Pani:2009wy}.
At the end of our study, we examined the stability of our solutions under axial perturbations. 

Our analysis began with an investigation of the properties of the spacetime geometry.
By expanding the metric components at infinity, we demonstrated that by appropriately choosing the integration constants, one can either obtain asymptotically flat or asymptotically (A)dS solutions.
In the context of the present work, we were solely focused on the study of asymptotically flat solutions, and we showed that the scalar curvature invariant quantities (the Ricci scalar $R=R^{\mu}{}_\mu$, the scalar $\mathcal{R}=R^{\mu\nu}R_{\mu\nu}$ and the Kretschmann scalar $\mathcal{K}=R^{\mu\nu\rho\sigma}R_{\mu\nu\rho\sigma}$) indicate the presence of a true spacetime singularity at $r=0$.
Examining then the causal structure of the spacetime, we found that the singularity is always surrounded by a horizon, implying that the derived solutions represent black holes. 
The resulting black-hole solutions have been classified into two categories, the \textit{normal} and the \textit{phantom} ones. 
The first class of solutions originate from regular/real scalar fields, while the second class from phantom scalar fields.
For any combination of the black-hole parameters $q$ and $M$ that leads to an apparent black-hole horizon, the horizon radius of a normal solution is always smaller than the horizon radius of the corresponding Schwarzschild black hole of the same mass. 
Conversely, a phantom solution with the same mass would have a greater horizon radius than that of a Schwarzschild black hole.
By appropriately selecting the scalar charge $q$ and the mass $M$, the normal and phantom solutions may lead to either ultra-compact or ultra-sparse black-hole solutions, respectively.
The former ones possess extremely low values of the ratio $r_h/(2M)$, while the latter have extremely high values of the same quantity. 
An immediate consequence of the above is that from a thermodynamic perspective and for a fixed black hole mass $M$, the ultra-sparse black holes appear to be the most stable solutions since their horizon entropy is greater than that of the corresponding Schwarzschild and ultra-compact solutions.

The fact that the black-hole solutions in the context of General Theory of Relativity are exclusively characterized by their mass, electromagnetic charge, and angular momentum, is a direct result of the no-hair theorems.
Similarly, no-scalar hair theorems \cite{NH-scalar1,NH-scalar2,Bekenstein} have been formulated for scalar-tensor theories of gravity, which forbid the association of black-hole solutions with scalar hair.
However, the no-scalar hair theorems are only viable in a subclass of the scalar-tensor theories. 
The theory we considered in this work constitutes a simple, yet special type of model that can evade the no-scalar hair theorem and lead to well-defined hairy black-hole solutions.
As it was analyzed in Sec. \ref{Sec: theory}, both normal and phantom solutions evade the no-scalar hair theorems for negative and positive-definite scalar potentials, respectively. Consequently, our hairy black-hole solutions are well-grounded. 

Having evaluated the analytic expressions of the scalar field and its potential, we then turned our attention to the components of the stress-energy tensor associated with the assumed scalar-field theory.
We showed that the normal solutions---which correspond to a regular/real scalar field---satisfy both null and strong energy conditions in the causal region of the spacetime, while in the interior of the black hole, the energy conditions are violated.
On the other hand, the phantom solutions violate the energy conditions in the causal region of spacetime.
This behavior though, is typical for phantom scalar fields, due to the fact that their kinetic terms come with the wrong sign.
This sole difference is sufficient to render the energy conditions violated.

Except for the examination of static solutions, we also investigated slowly rotating solutions.
The construction of slowly rotating black holes was achieved by treating the black-hole rotation as an axisymmetric perturbation on the background spherically symmetric metric.
The method we followed was first proposed by Hartle \cite{Hartle:1967he}, in the framework of General Relativity, and was then generalized by Pani and Cardoso for scalar-tensor theories \cite{Pani:2009wy}. 
We found that the angular velocity of our solutions is directly related to their scalar charge $q$.
As far as the normal solutions are concerned, we have shown that the more scalar charge they acquire, the greater their angular velocity becomes.
On the other hand, the phantom solutions exhibit the opposite behavior.
The greater their scalar charge $q$ is, the slower they rotate.
The reason behind this seemingly arbitrary connection lies in the fact that as the value of the scalar charge $q$ increases, the normal black-hole solutions become more compact, while the phantom black holes become more sparse.
As a result, between two rotating black hole solutions with the same physical characteristics, the one which is more compact is expected to rotate faster than the other one.
This behavior is indeed illustrated in Sec. \ref{Sec: rot-BHs}, and it also concurs with our physical intuition.

Finally, we have studied the stability of our solutions under axial (or odd) perturbations. 
We have successfully determined the Schr\"{o}dinger-like equation and the associated effective potential for our solutions.
Keeping in mind that the sole behavior of the effective potential is what decides the stability of a given solution, we have plotted the effective potential in terms of the dimensionless parameter $r^*/r_h$, for both normal and phantom black holes.
In the case of normal black-hole solutions with $\xi=5$, our analysis reveals that all resulting solutions are stable when $q/r_h$ is less than or approximately equal to 1.
On the other hand, for phantom black holes with $\xi=-2$, we have discovered that these solutions are stable for all values of $q/r_h$ within the range of $0$ to $\sqrt[4]{3}$.
Of course, a complete stability analysis demands the examination of polar (or even) perturbations as well. 
However, the study of even perturbations requires advanced mathematical methods, since it involves a system of four first-order differential equations with variable coefficients.
Hence, the sole analysis of even perturbations constitutes on its own a separate project.
Once the stability of a black-hole solution is fully analyzed, its quasi-normal modes (QNMs) can then be studied.
QNMs are the characteristic oscillations of a black hole that occur when it is perturbed by an external force or disturbance. 
These oscillations are damped and decay over time, and their frequencies depend only on the properties of the black hole, such as its mass, angular momentum, and charge, but not on the details of the perturbation.
Therefore, since different solutions are expected to possess different frequency spectra, QNMs constitute in a sense the identity of a compact object.
In conjunction with the fact that future updates of the LIGO-Virgo experiments might probe these frequencies, QNMs are likely to lead to the experimental verification of compact solutions originating from modified theories of gravity \cite{Berti:2018vdi}. 

Last but not least, the existence of hairy black holes in anti-de Sitter (AdS) spacetimes has attracted considerable attention due to its relevance to the AdS/CFT correspondence. This duality suggests a connection between gravity in AdS spacetimes and conformal field theories on their boundary. In this context, hairy black holes have been shown to play a crucial role in the phase structure and thermodynamics of the corresponding dual-field theory.

{\bf Acknowledgements.}
The research project was supported by the Hellenic Foundation for Research and Innovation (H.F.R.I.) under the “3rd Call for H.F.R.I. Research Projects to support Post-Doctoral Researchers” (Project Number: 7212).


\appendix

\section{The derivation of the scalar-field equation from the Einstein equations}
\label{App: sc-eq}
By expanding the $\nabla^\lam \nabla_\lam \Phi$ term in scalar-field equation \eqref{sc-eq} and multiplying the left hand side with $\Phi'$ we obtain
\gat$\label{appSE: sc1}
V'= B'\, \Phi'^2 + B\, \Phi'\, \Phi'' + \frac{B}{2r}(4+r\, A')\Phi'^2\,. $
In the above, we have used the fact that $V'=(\pa_\Phi V)\Phi'$. 
Using now \eqref{eq1} to substitute $\Phi'^2$ in terms of $A'$, we get
\eq$\label{appSE: sc2}
V'= \frac{2}{r}\, A'B'+ B\, \Phi'\, \Phi'' + \frac{BA'}{r^2}(4+r\, A')\,. $
Let us now focus on the gravitational field equations \eqref{eq2}-\eqref{eq3}. 
By differentiating both sides of eq. \eqref{eq3} with respect to $r$ and substituting the term $B''+B A''$ that will appear in the r.h.s. using eq. \eqref{eq2}, one should obtain \eqref{appSE: sc2}. 
This demonstrates the interdependence between the gravitational field equations and the scalar-field equation \eqref{sc-eq}, revealing that the latter can be derived from the former.

\section{Differential equation of the mass-function}
\label{App: DE}

\noindent The differential equation \eqref{eq2-new} is a second order, linear, and nonhomogeneous differential equation with variable coefficients, it is also presented below for convenience
\eq$\label{appDE: de}
B''(r) + \frac{3\xi q^2}{r\left(q^2+r^2\right)}\, B'(r) + 
2\, \frac{q^4(\xi^2-1)-2q^2r^2(\xi+1)-r^4}{r^2\left(q^2+r^2\right)^2}\, B(r) = -\frac{2}{r^2}\,.$
The general solution to the nonhomogeneous differential equation is given by the sum of the solution regarding the associated homogeneous differential equation, and a particular solution $B_p(r)$ of the nonhomogeneous one, namely
\eq$\label{appDE: gen-sol}
B(r) = B_h(r) + B_p(r)\,.$

The homogeneous differential equation
\eq$\label{appDE: hom-de}
B_h''(r) + P(r) B_h'(r) + Q(r) B_h(r) = 0\,, $
with 
\eq$\label{appDe: coef}
P(r)=\frac{3\xi q^2}{r\left(q^2+r^2\right)}\,, \hspace{1em} Q(r)=2\, \frac{q^4(\xi^2-1)-2q^2r^2(\xi+1)-r^4}{r^2\left(q^2+r^2\right)^2}\,,$
has the general solution
\eq$\label{appDE: Bh}
B_h(r) = \mathcal{A}_1\, B_1(r) + \mathcal{A}_2\, B_2(r)\,,$
where $\mathcal{A}_1$ and $\mathcal{A}_2$ are constants, while $B_1(r)$ and $B_2(r)$ are two linearly independent solutions. 
However, due to the fact that \eqref{appDE: hom-de} has variable coefficients we need to guess the first linearly independent solution $B_1(r)$, and then with the use of the Wronskian we can determine the second one. 
It is straightforward to verify that 
\eq$\label{appDE: B1sol}
B_1(r)=\left( r/q\right)^{2(1-\xi)} \left(1+r^2/q^2\right)^\xi$
is indeed a solution to \eqref{appDE: hom-de}. Using now the Wronskian
\eq$\label{appDE: wron}
W(r)=\left| \begin{array}{cc}
   B_1(r)  &  B_2(r)  \\
   B_1'(r) &  B_2'(r)
\end{array} \right| = B_1(r)B_2'(r)-B_2(r)B_1'(r)\,,$
and the fact that both $B_1(r)$ and $B_2(r)$ are solutions to the homogeneous equation, it is easy to show that
\eq$\label{appDE: wron-de}
\frac{W'(r)}{W(r)}=-P(r)\Leftrightarrow 
W(r) = W_0\, e^{-\int P(r)\, dr} = W_0\, (r/q)^{-3\xi}\left(1+r^2/q^2\right)^{3\xi/2} \,,$
where $W_0$ is an integration constant. 
Given that $B(r)$ is a dimensionless function, eq.\,\eqref{appDE: wron} indicates that the function $W(r)$ has dimensions of inverse length. 
Therefore, we may render the value of $W_0=1/q$. 
This particular choice provides the correct units to $W(r)$, while its numerical value will not affect the generality of the solution $B_h(r)$. 
Combining now eqs.\,\eqref{appDE: wron} and \eqref{appDE: wron-de} we obtain the differential equation regarding the function $B_2(r)$, 
that is
\eq$\label{appDE: B2-de}
B_2'(r)-\frac{B_1'(r)}{B_1(r)}\, B_2(r) = \frac{W(r)}{B_1(r)} =\frac{1}{q} \frac{e^{-\int P(r)\, dr}}{B_1(r)} \,.$
To solve the preceding equation we just need to multiply both sides with $[B_1(r)]^{-1}$.
By doing so, we obtain
\gat$\label{appDE: B2sol}
\frac{d}{dr}\left[\frac{B_2(r)}{B_1(r)} \right] =\frac{1}{q}  \frac{e^{-\int P(r)\, dr}}{[B_1(r)]^2} \Leftrightarrow 
B_2(r) = \frac{B_1(r)}{q} \int \frac{e^{-\int P(r)\, dr}}{[B_1(r)]^2}\, dr \,. $
In the above, we have ignored the integration constant since it does not add any additional information to the general solution
\eqref{appDE: Bh} of the homogeneous equation \eqref{appDE: hom-de}. 
Substituting the expressions of the functions $P(r)$ and $B_1(r)$ in \eqref{appDE: B2sol}, and after some algebra to make the integral dimensionless, one should get
\eq$\label{appDE: B2exp1}
B_2(r)= \left(r/q \right)^{2(1-\xi)}\left(1+r^2/q^2\right)^\xi \int \left( r/q \right)^{\xi-4} \left[ 1+\left( r/q \right)^2 \right]^{-\xi/2}\ d\left( r/q \right)\,.$
For $\xi\in \mathbb{Z}$ the above integral can be solved in terms of known functions, and thus $B_2(r)$ is determined analytically.
For example, one can verify that
\bal$
&\text{\underline{$\xi=-2$\,:}} &B_2(r)&=-\frac{1}{15}\,\frac{r}{q}\,\frac{ 3+5 r^2/q^2}{ \left(1+r^2/q^2\right)^2}\,,\nonum\\[2mm]
&\text{\underline{$\xi=-1$\,:}} &B_2(r)&=\frac{(r/q)^4\, \ln \left(1+\sqrt{1+r^2/q^2}\right)-(r/q)^4 \ln (r/q)-\left(2+r^2/q^2\right) \sqrt{1+r^2/q^2} }{
8 \left(1+r^2/q^2\right)}\,,\nonum\\[2mm]
&\text{\underline{$\xi=1$\,:}} &B_2(r)&=\frac{1}{2}\left(1+\frac{r^2}{q^2}\right) \left[\ln \left(1+\sqrt{1+r^2/q^2}\right)-\ln \left(r/q\right) -\frac{ \sqrt{1+r^2/q^2}}{r^2/q^2}\right]\,,\nonum\\[2mm]
&\text{\underline{$\xi=2$\,:}} &B_2(r)&=-\frac{\left(1+r^2/q^2\right)^2 }{r^3/q^3}\left[1+ (r/q) \arctan \left(r/q\right)\right]\,.\nonum$
However, for non-integer values of $\xi$, the best we can do is to express $B_2(r)$ in terms of the hypergeometric function $\,_2F_1(a,b;c;z)$ as presented below
\eq$\label{appDE: B2exp2}
B_2(r) = \left( r/q \right)^{-(1+\xi)} \frac{\left(1+r^2/q^2\right)^\xi}{\xi-3}\, \,_2F_1\left(\frac{\xi}{2},\frac{\xi-3}{2};\frac{\xi-1}{2};-\frac{r^2}{q^2}\right)\,. $
In the above, we have employed the integral representation of the hypergeometric function \cite{abramowitz+stegun}, namely
\eq$\label{appDE: hyper-int-rep}
\,_2F_1(a,b;c;z) = \frac{\Gamma(c)}{\Gamma(b)\Gamma(c-b)}\int_0^1 dt\ t^{b-1} (1-t)^{c-b-1} (1-z t)^{-a}\,,$
with $Re(c)>Re(b)>0$. 
Note also that an arbitrary constant should be introduced in \eqref{appDE: B2exp1} in order to convert the indefinite integral to a definite one.
This constant though does not add additional information to the final result, therefore it may ultimately be ignored.
Consequently, we can define the function $H(r)$ as follows
\eq$\label{appDE: H}
H(r) \equiv \left\{ \begin{array}{cr}
\displaystyle{\left(r/q\right)^{2(1-\xi)}\left(1+r^2/q^2\right)^\xi \int \left( r/q \right)^{\xi-4} \left[ 1+\left( r/q \right)^2 \right]^{-\xi/2}\ d\left( r/q \right)}\,, & \hspace{1em}\xi\in\mathbb{Z}\\[5mm]
\displaystyle{ \left( r/q \right)^{-(1+\xi)} \frac{\left(1+r^2/q^2\right)^\xi}{\xi-3}\, \,_2F_1\left(\frac{\xi}{2},\frac{\xi-3}{2};\frac{\xi-1}{2};-\frac{r^2}{q^2}\right)}\,, & \xi\notin\mathbb{Z}
\end{array}\right\}\,,$
which will henceforth be used interchangeably with the function $B_2(r)$.

Let us now determine the general solution $B(r)$ of the nonhomogeneous differential equation \eqref{appDE: de}.
To this end, we will use the method of variation of parameters or the Lagrange's method as it is often called.
The basic idea of the method is to replace the constants $\mathcal{A}_1$ and $\mathcal{A}_2$ of the homogeneous solution
\eqref{appDE: Bh} with the functions $\mathcal{A}_1(r)$ and $\mathcal{A}_2(r)$, respectively. 
Then, the general solution to the equation \eqref{appDE: de} will be of the form
\eq$\label{appDE: B}
B(r) = \mathcal{A}_1(r)B_1(r) + \mathcal{A}_2(r) B_2(r)\,, $
where $B_1(r)$ and $B_2(r)$ are given by the relations \eqref{appDE: B1sol} and \eqref{appDE: H}, respectively.
Substituting now the expression \eqref{appDE: B} in \eqref{appDE: de}, and after an appropriate grouping of terms, one
should obtain the following set of differential equations
\eq$\label{appDE: syst}
\left\{ \begin{array}{c}
\mathcal{A}_1'(r) B_1(r) + \mathcal{A}_2'(r) B_2(r) = 0  \\[2mm]
\displaystyle{\mathcal{A}_1'(r) B_1'(r) + \mathcal{A}_2'(r) B_2'(r) = -\frac{2}{r^2}}
\end{array}\right\}\,.$
It is straightforward to solve the above system and determine the unknown functions $\mathcal{A}_1(r)$, and $\mathcal{A}_2(r)$.
By doing so, we find that
\bal$\label{appDE: A1sol}
\mathcal{A}_1'(r) = \frac{2}{r^2}\, \frac{B_2(r)}{W(r)} \Leftrightarrow 
\mathcal{A}_1(r) &= C_1 + 2 \int (r/q)^{3\xi-2}\left(1+r^2/q^2\right)^{-3\xi/2} H(r)\, d(r/q)\,,\\[2mm]
\label{appDE: A2sol}
\mathcal{A}_2'(r) = -\frac{2}{r^2}\, \frac{B_1(r)}{W(r)}\Leftrightarrow 
\mathcal{A}_2(r) &= C_2 -2 \int (r/q)^\xi \left( 1+r^2/q^2 \right)^{-\xi/2} \, d(r/q) \,.$
%

Using now \eqref{appDE: B} and substituting the expressions of the functions $B_1(r)$, $B_2(r)$, $\mathcal{A}_1(r)$ , 
and $\mathcal{A}_2(r)$ as they are given by \eqref{appDE: B1sol}, \eqref{appDE: H}, \eqref{appDE: A1sol}, 
and \eqref{appDE: A2sol}, respectively, we get the solution
\bal$B(r) = &\,\left( r/q\right)^{2(1-\xi)} \left(1+r^2/q^2\right)^\xi \left[ C_1 + 2 \int (r/q)^{3\xi-2}\left(1+r^2/q^2\right)^{-3\xi/2} H(r)\, d(r/q)\right] \nonum\\
& + H(r) \left[ C_2 -2 \int (r/q)^\xi \left( 1+r^2/q^2 \right)^{-\xi/2} \, d(r/q) \right] \,, $
which for non-integer values of $\xi$ can be brought to the following form
\bal$\label{appDE: B-r-xi-non-int}
B(r) = &\, C_1\, (r/q)^{2(1-\xi)} \left( 1+r^2/q^2 \right)^\xi + \frac{C_2}{\xi-3}\, (r/q)^{-(1+\xi)} \left( 1+r^2/q^2 \right)^\xi 
\,_2F_1 \left( \frac{\xi}{2}, \frac{\xi-3}{2}; \frac{\xi-1}{2}; -\frac{r^2}{q^2} \right) \nonum\\
& + \frac{2 \left( 1+r^2/q^2 \right)^\xi}{\xi-3} \Bigg\{
 \left(\frac{r}{q}\right)^{2(1-\xi)} \int \,_2F_1 \left( \frac{\xi}{2}, \frac{\xi-3}{2}; \frac{\xi-1}{2}; -\frac{r^2}{q^2} \right) \left(\frac{r}{q}\right)^{2\xi-3} 
\left( 1+\frac{r^2}{q^2} \right)^{-\xi/2} d\left(\frac{r}{q}\right)\nonum\\
& - \frac{1}{\xi+1}\, \,_2F_1 \left( \frac{\xi}{2}, \frac{\xi-3}{2}; \frac{\xi-1}{2}; -\frac{r^2}{q^2} \right)
\,_2F_1 \left( \frac{\xi}{2}, \frac{\xi+1}{2}; \frac{\xi+3}{2}; -\frac{r^2}{q^2} \right) \Bigg\} \,.$
In the preceding expression, the identity \eqref{appDE: hyper-int-rep} has been employed.



\section{Additional asymptotically flat solutions}
\label{App: Add-sol}

\noindent Due to their interesting expressions, we present here the functions $B(r)$ for $\xi=1,2$. Hence, for $\xi=1$ one can verify that
\eq$
B(r)=\left(1+\frac{r^2}{q^2}\right) \left[\frac{q^2}{r^2} - \frac{3M}{q}\frac{q^2}{r^2}\sqrt{1+\frac{r^2}{q^2}} + \frac{3M}{q}\,\ln \left(\frac{1+\sqrt{1+r^2/q^2}}{r/q}\right)\right]\,,$
while for $\xi=2$,
\bal$
B(r)=\left(1+\frac{r^2}{q^2} \right)^2 \Bigg\{ \frac{3M}{r}\left(\frac{\pi q}{r}-\frac{2q^2}{r^2} \right) - \frac{\pi q^2}{r^2}\left( \frac{\pi}{4} - 
\frac{q}{r}\right) - \left( \frac{6Mq}{r^2} + \frac{2q^3}{r^3} - \frac{\pi q^2}{r^2} \right) \tan^{-1}\left( \frac{r}{q}\right) \nonum\\[1mm]
- \frac{q^2}{r^2}\, \left[\tan^{-1}\left( \frac{r}{q}\right) \right]^2 \Bigg\}\,.$


\section{The analytic formulas for the curvature invariants}
\label{App: Curv-inv}

The analytic formulas for the curvature invariant quantities $R\equiv R^{\mu}{}_{\nu}$, $\mathcal{R}\equiv R^{\mu\nu}R_{\mu\nu}$, and $\mathcal{K}\equiv R^{\mu\nu\rho\sigma}R_{\mu\nu\rho\sigma}$, obtained from the line-element \eqref{metr-ans}, are presented below.

For the normal solution with $\xi=5$, one finds that
\bal$\label{appCurv: R-real}
R&=\frac{2q^2}{3\,r^{10}}\left(15\,r^6+100\,q^2 r^4+130\,q^4 r^2+51\,q^6 \right)-\frac{108 M q^7}{3\,r^{10}}\left(1+\frac{r^2}{q^2} \right)^{5/2}\,,\\[5mm]
\label{appCurv: Ric-real}
\mathcal{R}&=\frac{4 q^4}{27\,r^{20}}\big( 675\, r^{12} + 6300\, q^2r^{10} + 23475\, q^4 r^{8} + 42360\, q^6 r^{6} + 40075\, q^{8} r^{4} + 19295\, q^{10} r^{2} + 3757\, q^{12} \big)\nonum\\[1mm] 
& \hspace{1.5em}  + \frac{624\, M^2 q^{14}}{r^{20}} \left(1+\frac{r^2}{q^2} \right)^5 - \frac{8\, M q^9}{3\, r^{20}}\left(1+\frac{r^2}{q^2}\right)^{5/2}\big(180\, r^{6} + 900\, q^2 r^{4}  + 1135\,q^{4} r^{2} + 442\, q^{6} \big)\,,\\[5mm]
\label{appCurv: Rim-real}
\mathcal{K}&= \frac{4 q^4}{27\,r^{20}}\big( 2025\, r^{12} + 12000\, q^2r^{10} + 35610\, q^4 r^{8} + 58470\, q^6 r^{6} + 53306\, q^{8} r^{4} + 25330\, q^{10} r^{2} + 4913\, q^{12} \big)\nonum\\[1mm] 
& \hspace{1.5em} + \frac{48\, M^2 q^{14}}{r^{20}} \left( 17+ \frac{r^4}{q^4} \right) \left(1+\frac{r^2}{q^2} \right)^5 - \frac{16\, M q^7}{3\, r^{20}}\left(1+\frac{r^2}{q^2}\right)^{5/2}\big(30\, r^{8} + 175\, q^2 r^{6} + 616\, q^4 r^{4} \nonum\\[1mm]
& \hspace{1.5em} + 745\,q^{6} r^{2} + 289\, q^{8} \big)\,,$
while in the case of the phantom solution, where $\xi=-2$, one obtains
\bal$\label{appCurv: R-phant}
R_\mathfrak{p}&= \frac{4}{5\, q^4 r^2} \left(1+\frac{r^2}{q^2}\right)^{-3} \big[ (18M-5r)\, r^3 +2\, q^2(3M+5r)\, r + 5q^4 \big] \,,\\[5mm]
\label{appCurv: Ric-phant}
\mathcal{R}_\mathfrak{p}&= \frac{16}{75\,q^8 r^4} \left( 1+\frac{r^2}{q^2}\right)^{-6} \big[ 3(156 M^2 - 120 M r +25\, r^2)\, r^6 + 3 q^2(144 M^2 + 30 M r -25\, r^2)\, r^4 \nonum\\[1mm]
& \hspace{1.5em} + 12 M q^4 (9M+25\,r)\, r^2 +15\, q^6(6M+5\, r)\, r +25\, q^8 \big]  \,,\\[5mm]
\label{appCurv: Rim-phant}
\mathcal{K}_\mathfrak{p}&= \frac{16}{75\,q^{12}\, r^4} \left( 1+\frac{r^2}{q^2}\right)^{-6} \big[ 225 M^2 r^{10} + 300 M q^2(3M+r)\, r^8 + 3q^4(654 M^2 + 220 M r +75\, r^2)\, r^6 \nonum\\[1mm]
& \hspace{1.5em} + 2 q^6(774 M^2 + 840 M r + 175\, r^2)\, r^4  + 5 q^8 (81 M^2 +300M r+ 110\,r^2)\, r^2 \nonum\\[1mm]
& \hspace{1.5em} + 30\, q^{10} (14M+15\, r)\, r + 125\, q^{12} \big] \,.$
%


\addcontentsline{toc}{section}{References}
\bibliography{Bibliography}{}
\bibliographystyle{utphys}

\end{document}